\documentclass[num-refs]{wiley-article}

\usepackage{siunitx}
\usepackage{graphicx}
\usepackage{subcaption}
\usepackage{rotating}
\usepackage{booktabs}
\usepackage{amsfonts}
\usepackage{amssymb}
\usepackage{mathtools}
\usepackage{newfloat}
\usepackage{threeparttable}
\usepackage{url}

\usepackage{pifont}
\newcommand{\cmark}{\ding{61}}
\newcommand{\dmark}{\ding{117}}

\usepackage{amsmath}
\DeclareMathOperator*{\argmax}{argmax} 

\papertype{Original Article}
\paperfield{Journal Section}

\title{Active learning for efficiently training emulators of computationally expensive mathematical models}

\abbrevs{ALC|ALM: Active learning with the Cohn | Mackay criterion; BTGP: Bayesian treed Gaussian Process; GP: Gaussian Process; LHS: Latin hypercube sampling; PSA: Prostate-specific Antigen; QAD: Quality adjusted day.}

\author[1,2]{Alexandra~G.~Ellis}
\author[1]{Rowan Iskandar}
\author[1,3]{Christopher H.~Schmid}
\author[4]{John~B.~Wong}
\author[1]{Thomas~A.~Trikalinos}

\affil[1]{Center for Evidence Synthesis in Health, Brown University School of Public Health, Providence, RI 02912, US}
\affil[2]{Stratevi, Boston, MA 02111, US}
\affil[3]{Department of Biostatistics, Brown University School of Public Health, Providence, RI 02912, US}
\affil[4]{Division of Clinical Decision Making, Tufts Medical Center, Boston, MA 02111, US}

\corraddress{AG Ellis, Center for Evidence Synthesis in Health, Brown University School of Public Health, Providence, RI 02912, US}
\corremail{alexandra\_g\_ellis@brown.edu}

\fundinginfo{AG Ellis was supported by a dissertation research grant from the Agency for Healthcare Research and Quality. Grant Number: R36HS24653-01}

\runningauthor{Ellis \emph{et al}.}

\begin{document}
\maketitle

\begin{abstract} 
{\small
An emulator is a fast-to-evaluate statistical approximation of a detailed mathematical model (simulator). When used in lieu of simulators, emulators can expedite tasks that require many repeated evaluations, such as sensitivity analyses, policy optimization, model calibration, and value-of-information analyses. Emulators are developed using the output of simulators at specific input values (design points). Developing an emulator that closely approximates the simulator can require many design points, which becomes computationally expensive. We describe a self-terminating active learning algorithm to efficiently develop emulators tailored to a specific emulation task, and compare it with algorithms that optimize geometric criteria (random latin hypercube sampling and maximum projection designs) and other active learning algorithms (treed Gaussian Processes that optimize typical active learning criteria). We compared the algorithms' root mean square error (RMSE) and maximum absolute deviation from the simulator (MAX) for seven benchmark functions and in a prostate cancer screening model. In the empirical analyses, in simulators with greatly-varying smoothness over the input domain, active learning algorithms resulted in emulators with smaller RMSE and MAX for the same number of design points. In all other cases, all algorithms performed comparably. The proposed algorithm attained satisfactory performance in all analyses, had smaller variability than the treed Gaussian Processes (it is deterministic), and, on average, had similar or better performance as the treed Gaussian Processes in 6 out of 7 benchmark functions and in the prostate cancer model.
}

\end{abstract}

\section{Introduction}\label{sec:intro}
Many decisions in health must be made under uncertainty or with incomplete understanding of the underlying phenomena. In such cases, mathematical modeling can help decision-makers synthesize evidence from different sources, estimate and aggregate anticipated outcomes while accounting for stakeholder preferences, understand trade-offs, and quantify the impact of uncertainty on decision making \cite{Owens2016}. To be informative, a model should be detailed enough to capture salient aspects of the decisional problem at hand, but a highly detailed model can render routine analyses computationally expensive, hindering its usability. 

As an example, consider the prostate-specific antigen (PSA) growth and prostate cancer progression (PSAPC) model, which is a microsimulation model that projects outcomes for PSA-based screening strategies defined by combinations of screening schedules and age-specific PSA positivity thresholds~\cite{Gulati2010, Gulati2013}. 
Although evaluating the expected outcomes of one screening strategy takes just a few minutes on modern computer hardware, analyses that require many (on the order of $10^4$ to $10^6$) model evaluations, such as \emph{calibration} of model parameters to external data~\cite{NRC2012,rutter2009}, \emph{policy optimization}~\cite{Jones1998ego}, \emph{sensitivity analysis}~\cite{NRC2012,saltelli2008,Kleijnen2008}, and \emph{uncertainty analysis}~\cite{NRC2012,OHagan2006,decarvalho2019}, become expensive if not impractical. 
For example, in part because of the computational burden, Gulati and colleagues evaluated only $35$ screening strategies~\cite{Gulati2013}, out of more than $10^7$ policy-relevant strategies that could have been explored~\cite{bertsimas2018}.

A way to mitigate the computational cost of such analyses is to use \emph{emulators} of the original mathematical models. Emulators, also called surrogate models or meta-models, are statistical approximations of the original models, hereafter referred to as \emph{simulators} for consistency with prior works~\cite{Kleijnen2008,OHagan2006,santner2003,Sacks1989}. 
The goal of the emulation depends on the task at hand: the emulator may approximate a simulator over the whole input domain (e.g., to expedite sensitivity, uncertainty, or value-of-information analyses), or in the neighborhood of a critical set of inputs (e.g., for optimization tasks, in the neighborhood of an optimum).
Once developed, emulators are orders of magnitude faster to evaluate than the simulators and, thus, can be used instead of, or in tandem with, the simulators to perform computationally intensive tasks~\cite{Kleijnen2008,Blanning1975,Sacks1989}. 
Although uncommon in healthcare applications~\cite{Neumann2018,Jalal2015,neumann2016}, using emulators in place of simulators is a well-established practice in mechanical, electrical, and industrial engineering, geology, meteorology, and military modeling \cite{Kleijnen2008, Conti2010,Kennedy2006, NRC2012, OHagan2006}. 

Developing an emulator that approximates a simulator over an input subdomain is analogous to estimating a response surface, so it is an experimental-design problem~\cite{Kleijnen2004}. In this context, the simulator is used to generate data (a \emph{design}), and an emulator is fit to the design. We describe an adaptive algorithm for developing designs that are tailored to a specific simulator and a particular emulation task. We compare the performance of emulators generated with the adaptive design versus alternative adaptive and non-adaptive designs using benchmark functions and the PSAPC model as test cases.

This work is organized as follows. In Section~\ref{sec:emulation}, we describe desirable characteristics for emulators of simulators in health to motivate our choice of Gaussian Process emulators. In Section~\ref{sec:algorithm}, we review common designs for developing emulators and introduce another adaptive design, which is the focus of this work. In Section~\ref{sec:examples} we describe the experiments for comparing our algorithm with emulator-free and alternative emulator-based algorithms.
We first explore the behavior and performance of our various designs under different situations using benchmark functions (Section~\ref{sec:benchmarks}), and then apply these algorithms to emulate the PSAPC model to develop emulators that predict gains in life expectancy over no screening for large sets of practically-implementable prostate cancer screening strategies (Section~\ref{sec:PSAPC_model}). Although our adaptive design is meant for very expensive simulators (that, e.g., take hours or days rather than minutes to evaluate), we use the PSAPC simulator as a model with which we can actually do full computations. We present results from the experiments in Section~\ref{sec:results} and conclude with key remarks in Section~\ref{sec:remarks}.

\section{Simulators, Emulation goals, and emulator models}\label{sec:emulation}

\subsection{Simulators}\label{sec:simulators}

For this exposition, a simulator is a deterministic smooth function $f: \mathbb{R}^K \rightarrow \mathbb{R} $ that maps from $K$ input parameters to scalar outputs. $K$ is typically in the many dozens or hundreds. We treat the simulator as a \emph{black box} function that can be evaluated (expensively) at specific input values. Typically, a baseline set of values $\bm{z}^* \in \mathbb{R}^K$ has been specified for the simulator through data analysis, calibration, or expert knowledge. 

While this setup is not the most general case for which we could present our algorithm, it is not as restrictive as it appears for the following reasons: (i) Many simulators in health that are not smooth are Lipschitz-continuous (i.e. have no ``extreme'' jumps in the output given a marginal change in input) and, for the precision demanded in practice and for our purposes, can be treated as if they were smooth functions. (ii) Simulator outputs can be random variables, e.g., if some inputs are random variables, but we are usually interested in expectations of outcomes, which average over the random input variables and result in deterministic mappings. (iii) Most simulators have multidimensional outputs, but, often, we are interested in one critical or composite outcome or are willing to consider one outcome at a time. (iv) Finally, we treat the simulator as a black box and do not attempt to exploit its analytical form. Often the analytical form is not available, or, when it is, its theoretical analysis may severely test our mathematical ability or be intractable. In Section~\ref{sec:remarks} we discuss potential extensions of our work to stochastic simulators and to simulators with multidimensional outputs.

\subsection{Goal of emulation}\label{sec:emulation_goals}

We aim to develop an emulator $f^*(\cdot)$ that statistically approximates the simulator's input-output mapping over the domain $\mathcal{X} \subseteq \mathbb{R}^k$ of the simulators' input parameters, where ${1 \le k \le K}$~\cite{OHagan2006}. We seek 
\begin{equation}\label{eq:def_emulator}
    f^*(\bm{x}) \cong f(\bm{x}, \bm{w}^*)  \ \forall \ \bm{x} \in \mathcal{X},
\end{equation}
\noindent where $\bm{x}$ are values for the $k$ inputs whose mapping we wish to emulate, $\bm{w}^* \in \mathbb{R}^{K-k}$ are values for the remaining inputs of the simulator that are kept equal to their corresponding elements in the baseline-values vector $\bm{z}^*$, and the symbol ``$\cong$'' means ``either equal or close enough for the purpose of the application''. Hereafter, we write  $f(\bm{x})$ instead of $f(\bm{x}, \bm{w}^*)$ to ease notation. We assume that $\mathcal{X}$ is a \emph{polytope}, i.e., a convex polyhedron in $\mathbb{R}^k$, which is most often the case in applications. 

Approximating the behavior of the emulator \emph{over all vectors} in $\mathcal{X}$ is a reasonable goal when we wish to gain insights about the behavior of the simulator or for sensitivity and uncertainty analysis. For different tasks, e.g., for calibration of input variables to external data, it may suffice to approximate the simulator in the \emph{neighborhood of the set of optimal values}. We describe algorithms tailored to finding optima elsewhere~\cite{ellis2018}.  

To fit $f^*(\cdot)$, we need a design that specifies the set of $n$ points in $\mathcal{X}$ at which we will evaluate the simulator. Let $\mathcal{X}_n = \{ \bm{x}_1, \ldots, \bm{x}_n \}$ be a set of distinct vectors in $\mathcal{X}$ that includes the extreme vectors of $\mathcal{X}$, so that  all vectors in $\mathcal{X}$ are convex combinations of vectors in $\mathcal{X}_n$. Then, with some abuse of terminology, we will call the vectors in ${\mathcal{D}_n = \{ (\bm{x}^T, f(\bm{x}))^T : \bm{x} \in \mathcal{X}_n \}}$ the \emph{design vectors} or \emph{design points}. We further simplify notation by writing ${(\bm{x}, f(\bm{x}))}$ instead of ${(\bm{x}^T, f(\bm{x}))^T}$ for each vector in $\mathcal{D}_n$ .

\subsection{Choice of emulator model} \label{sec:emulatormods}

We follow others in requiring that the emulator be an \emph{exact interpolator}~\cite{ohagan1998,NRC2012,Kleijnen2008,OHagan2006,Sacks1989}. Specifically, we demand that  
\begin{enumerate}
\item $f^*(\bm{x})=f(\bm{x})$ for $\bm{x} \in \mathcal{X}_n$, that is, the emulator’s prediction should agree with the simulator’s output value at the design points because the simulator is deterministic.\label{item:exact}
\item At all other input vectors, the emulator must provide an interpolation of the simulator output value, whose uncertainty decreases when closer to a design point, so that it becomes 0 at the design points.  \label{item:interpolator}
\item The emulator should be orders of magnitude faster to evaluate than the simulator.\label{item:fast}
\end{enumerate}

\begin{figure}[hbt!]
\centering
\begin{subfigure}[t]{.48\textwidth}
\includegraphics[width=\textwidth]{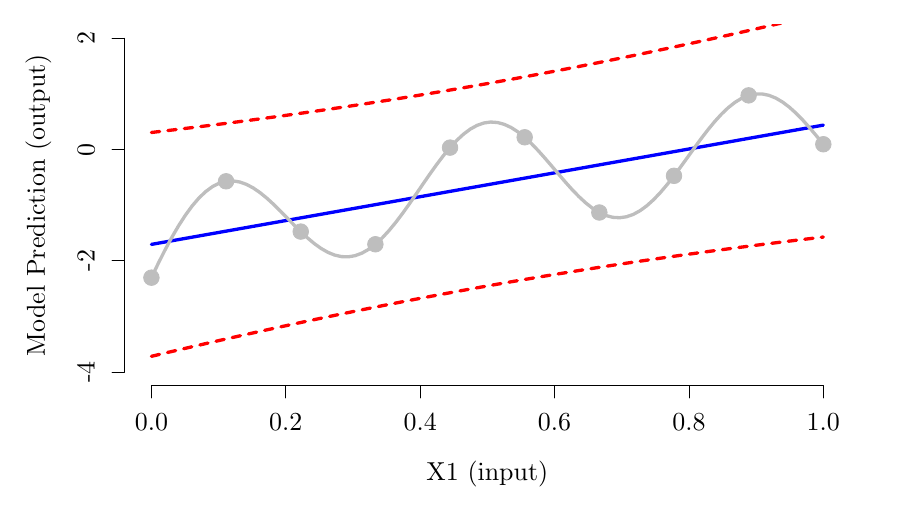}
\subcaption{Regression emulator.}
\end{subfigure}
\begin{subfigure}[t]{.48\textwidth}
\includegraphics[width=\textwidth]{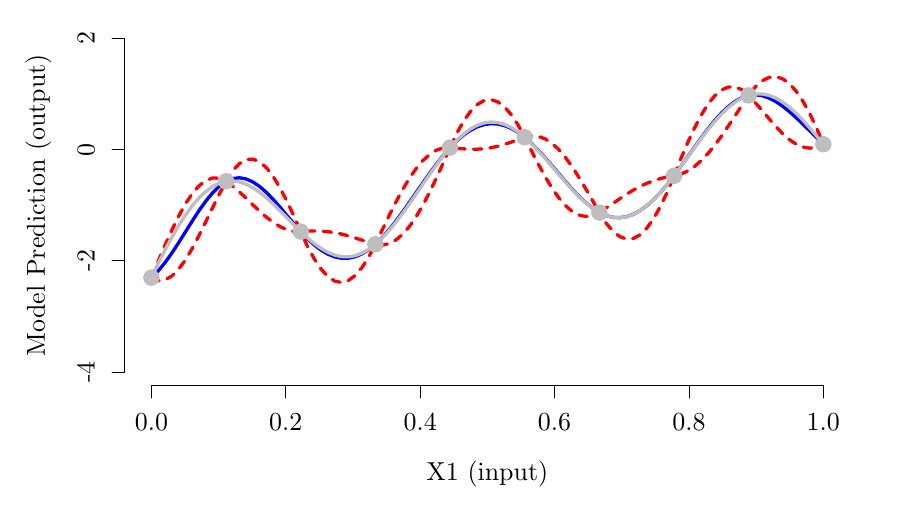}
\subcaption{Gaussian Process emulator.}
\end{subfigure}
\caption[]{\textbf{Two emulator models.} Two emulator functions predict the output of a simplistic model (grey curve). The input is a single variable (X1, horizontal axis). The 10 grey points represent the data (``design points'') used to fit the emulators. The solid blue curves are the emulators' predictions, and the areas between the red dashed curves are the associated 95\% prediction intervals. The prediction from the linear regression emulator (a) passes near, but not through, all design points. Further, the prediction uncertainty is not zero at the design points. The Gaussian Process emulator (b) passes through all design points, and the uncertainty of its predictions has the desired pattern. Figure adapted from~\cite{NRC2012}. 
\label{fig:GPexample}}
\end{figure}

Despite the fact that practically all statistical and machine learning models are fast to evaluate, i.e., they satisfy criterion~\ref{item:fast}, typical approaches such as linear regressions and neural networks do not satisfy the first two criteria, whereas kernel-based methods such as Gaussian Processes (GPs), do. Figure~\ref{fig:GPexample} helps build intuition. Despite its popularity in the literature in part due to its simple form and familiarity by researchers \cite{Batmaz2003, Kleijnen2008}, regression is not a preferred emulator type in this context. We use GP-based emulation, also referred to as \emph{kriging} in other fields~\cite{NRC2012}.

Depending on prior knowledge about the simulator or the analysis at hand, one may choose between GPs that assume the observations (design points) have no noise (for deterministic simulators) or are noisy (for stochastic simulators), and between stationary versus various types of non-stationary GPs. Gramacy and Lee 2012 offer practical and philosophical arguments for defaulting to GPs that allow for noisy observations even when simulators are deterministic~\cite{gramacy2012}. This practice is akin to fitting regularized GP models~\cite{Rasmussen2006} and  has numerical advantages for GP parameter estimation~\cite{ranjan2011}, but can affect the shape and modes of the GP likelihood function~\cite{pepelyshev2010} and can result in fitted GPs that do not fulfill the criteria~\ref{item:exact} and~\ref{item:interpolator} above.   
We review stationary GPs briefly in Appendix~\ref{app:GPs}.

\subsection{Experimental designs for developing emulators}

We categorize experimental designs into \emph{emulator-free} and \emph{emulator-based} designs. Emulator-free designs optimize geometric criteria over $\mathcal{X}$ without leveraging any information about the simulator's mapping. In emulator-based design algorithms, points are selected sequentially: The next design points are selected by exploiting information from an emulator learned on the thus-far observed design points. These approaches typically require fewer design points than emulator-free designs \cite{Kleijnen2008,Park2002}. 

\subsubsection{Emulator-free designs}\label{sec:emulator-free_designs}
Typical emulator-free designs optimize criteria based on distance metrics, such as the Euclidean distance between any two input vectors. The number $n$ of design points (size of the design) is specified \emph{ex ante} based on prior understanding of the simulator's behavior or is limited by a computational budget so that $n \le B$. 
Optimizing for different geometric criteria results in designs with different properties. For example, the miniMax design selects $\mathcal{X}_n$ so that the maximum distance between any vector in $\mathcal{X}$ and a vector in $\mathcal{X}_n$ is minimal, and the Maximin design selects $\mathcal{X}_n$ so that the minimum distance between any two vectors in $\mathcal{X}_n$ is maximal~\cite{johnson1990}. By construction, the miniMax and Maximin designs have space-filling properties in the $k$ dimensions of $\mathcal{X}$, but they are generally not space-filling when projected on lower-dimensional $(1, 2, \dots, k-1)$ subspaces of $\mathcal{X}$~\cite{pronzato2012}. 
On the other hand, Latin Hypercube Designs (LHDs) select vectors that have space-filling properties in each input dimension, but not necessarily in higher ($2, \dots, k$)-dimensional subspaces of $\mathcal{X}$. Maximum projection (MaxPro) designs are analogous to miniMax, but retain space-filling properties when projected to any lower dimensional subspace of $\mathcal{X}$~\cite{joseph2015}.

In our empirical analyses, we examine designs sizes $n \le B=100$. We use Latin Hypercube Sampling (LHS), a LHD design which involves uniformly random sampling over prespecified partitions of each input dimension~\cite{mckay1979}, and MaxPro designs, because of their attractive space-filling properties when projected to all lower-dimensional subspaces of $\mathcal{X}$.

\subsubsection{Emulator-based design algorithms}\label{sec:emulator-based_designs}
Emulator-based designs are sequential or active learning~\cite{pronzato2012,cohn1994,mackay1992} in that they iteratively augment a seeding design by leveraging information about the simulator's mapping from a surrogate model (here, a GP-based emulator). In each iteration, new design points are selected, the emulator is updated (re-fit) to include the additional points, and the process continues until a stopping criterion is met. Thus, the type of emulator chosen is a part of the active learning algorithm.  Many active learning design algorithms have been proposed in the literature~\cite{pronzato2012,cohn1994,mackay1992, Kleijnen2008}, and we put forward yet another one in Section~\ref{sec:algorithm}. To help contextualize our contribution, we emphasize three 
high-level attributes of active learning design algorithms:    

\begin{enumerate}
    \item \emph{Criteria used to select the next design point (active learning criteria):} Design points may be chosen based on any criterion related to an emulator's fit or uncertainty. A natural approach is to choose a design point that optimizes a function of the emulator's predictive variance~\cite{mackay1992}. One option is to select the design point that minimizes the integrated variance of the emulator predictions over $\mathcal{X}$, as per Cohn 1994~\cite{cohn1994}; we will refer to active learning with the Cohn criterion as ALC. Another option is to select the point at which the prediction variance of the emulator is largest, as per MacKay 1992~\cite{mackay1992}; we will refer to active learning with the MacKay criterion as ALM. For a deeper motivation of these and related criteria, see~\cite{cohn1994, mackay1992} and Chapter 7.2 in Fedorov 1972~\cite{fedorov1972}. 
    \newline
    Intuitively, if the emulator is biased  (e.g., ``overfit'') early on, the resultant design may be adversely affected. To help mitigate the impact of a potentially biased emulator, in Section~\ref{sec:algorithm} we propose an active learning criterion that ascertains active learning criteria over \emph{a set of jack-knifed emulators}.  
    %
    \item \emph{Approach to optimizing the active learning criteria:} Optimizing the aforementioned criteria can be a formidable problem itself because of local optima. For example, in Section~\ref{sec:emulatormods}, we required that the emulator's uncertainty must decrease when closer to a design point so that it is 0 at the design point, which creates multiple local optima for the emulator's predictive variance. We find the global optimum for the active learning criteria in one of two ways, depending on what is more practical in the corresponding computational setting: (i) we optimize criteria over a discretized subset $\ \mathcal{X}_N \subset \mathcal{X}$ comprising $N>B$ input points from a space-filling design or from uniform sampling, or (ii) we find the global optimum using the partitioning and local searching algorithm described in Section~\ref{sec:Algorithm2}. 
    \item \emph{Specification of GP-based emulator,} which depends on prior knowledge about the simulator or the analysis at hand.  
\end{enumerate}

Many sequential design algorithms can be constructed to address different challenges. For example, for a deterministic simulator with reasonably constant smoothness throughout the input domain (such as those described in Section~\ref{sec:simulators}), active learning algorithms based on stationary GPs, where the next point is selected by the ALM~\cite{mackay1992} or ALC~\cite{cohn1994} criteria are reasonable choices. In section~\ref{sec:benchmarks} we explore several such functions with 1, 2, and 3 input dimensions.

A more challenging situation arises when the simulator's smoothness varies substantially over the input domain. Then the mean of the observations, their variance (for noisy-observations from stochastic simulators), or the covariance function of the GP may depend on the inputs and a stationary GP would not fit well. Instead, one may favor  GPs that model the mean~\cite{OHagan2006,OHagan2006b,Rasmussen2006} or variance of observations~\cite{binois2018} as functions of the inputs, or GPs that use covariance functions other than the squared exponential kernel we used here (see Rasmussen 2006 for a review ~\cite{Rasmussen2006}). Alternatively, one may fit stationary GPs in different partitions of the input space: For example, Gramacy and Lee 2008 proposed a treed-GP emulator model that uses a divide and conquer strategy. Their model learns an axis-aligned partitioning of the input space so that the simulator has  relatively constant smoothness in each partition and then fits a stationary GP in each partition~\cite{gramacy2008}.  In Section~\ref{sec:benchmarks} we use such a benchmark function, where the mean changes substantially over the input domain.

To compare our proposed algorithm against state-of-the-art active learning algorithms, we implement two active learning algorithms that use the treed GP emulator model of Gramacy and Lee and the ALC and ALM criteria to select the next design points~\cite{gramacy2008}. During active learning with treed GPs, we examined whether an alternative partitioning of the input space fit the data better every 10 additional design points. We also required that a partition includes at least 10 design points.

\section{An active learning algorithm} \label{sec:algorithm}
\begin{figure}
\captionsetup[subfigure]{margin={0cm,0cm}}
   \parbox[b][2in][t]{0.33\textwidth}{\includegraphics[width=.9\hsize]{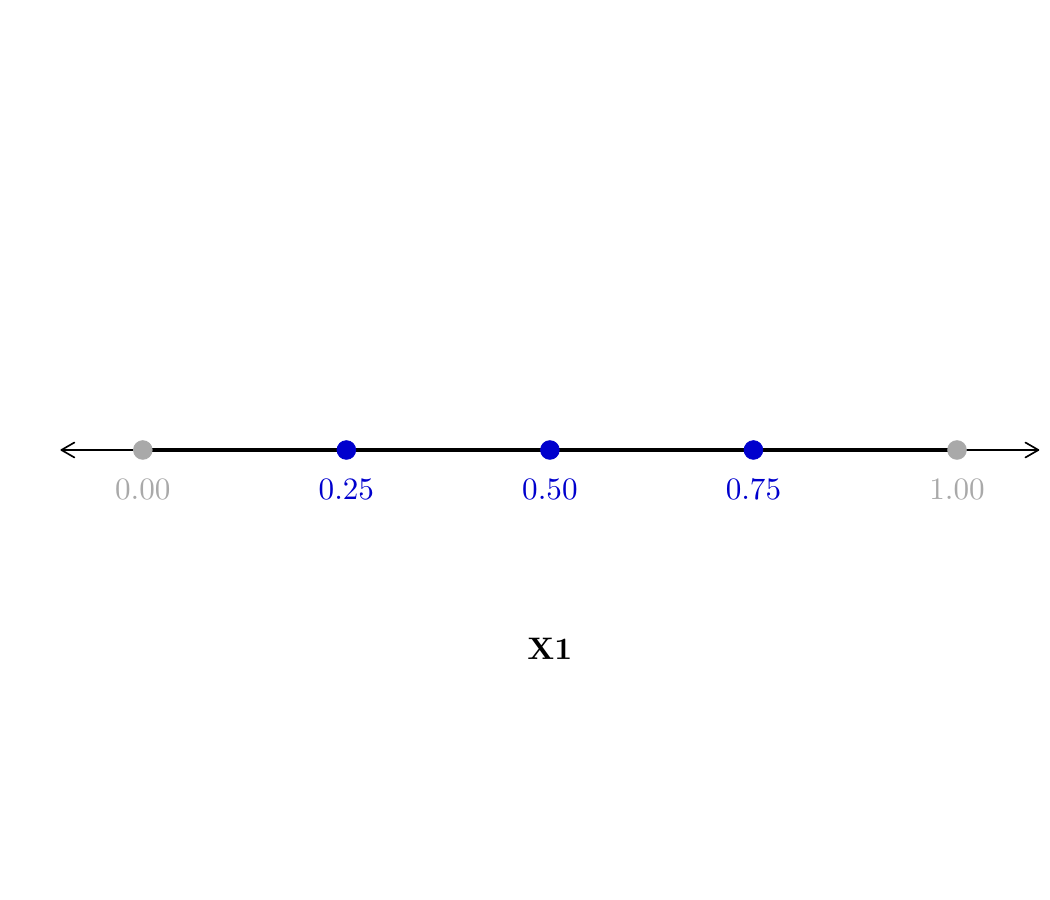} \subcaption{Initial set-up: Input vectors.}}%
   \parbox[b][2in][t]{0.33\textwidth}{\includegraphics[width=.9\hsize]{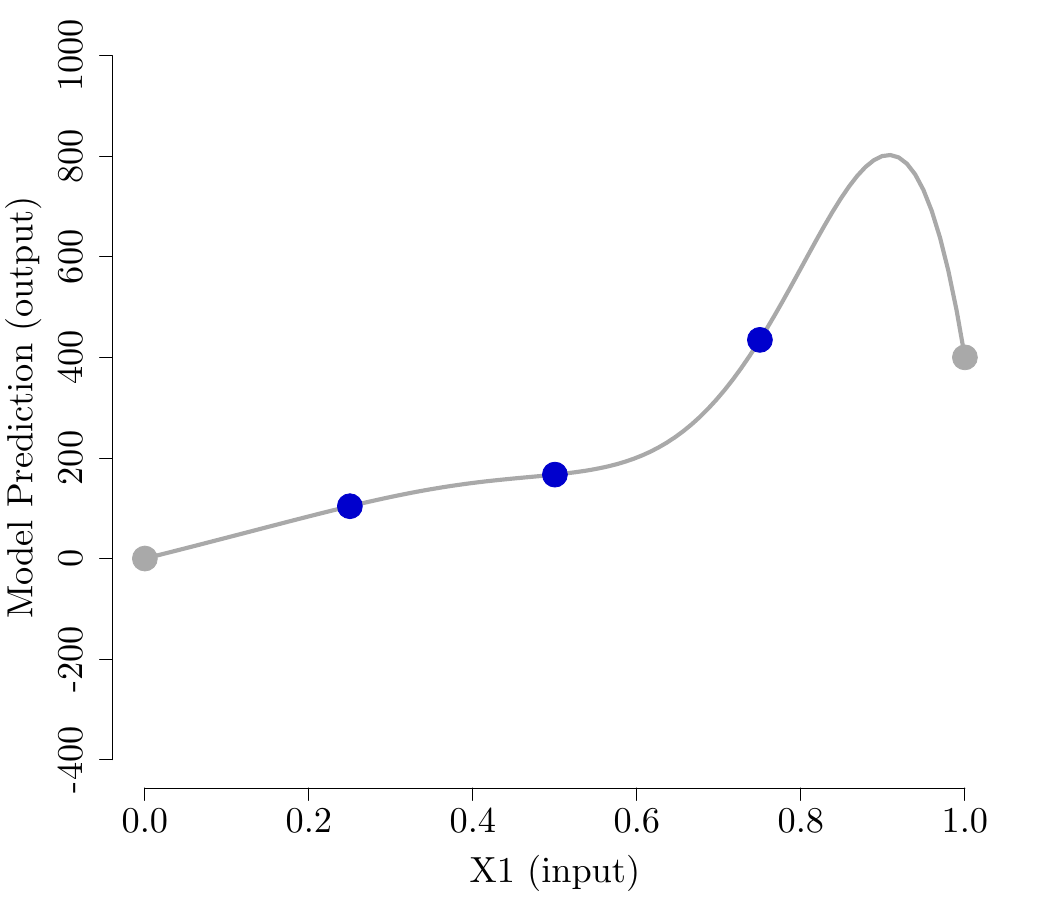} \subcaption{Initial set-up: Design points.}}%
   \parbox[b][2in][t]{0.33\textwidth}{\includegraphics[width=.9\hsize]{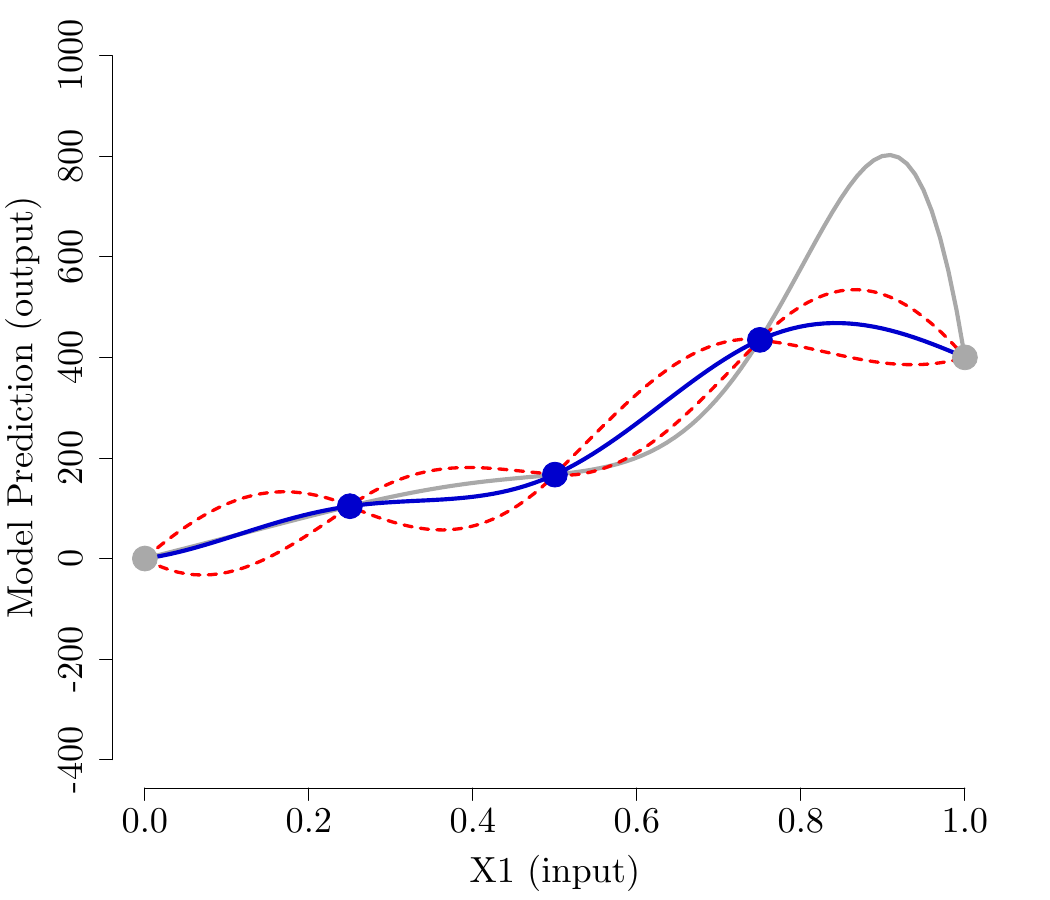} \subcaption{Fit an emulator (Step 0).}}\\[1ex]
   \parbox[b][2in][t]{0.33\textwidth}{\includegraphics[width=.9\hsize]{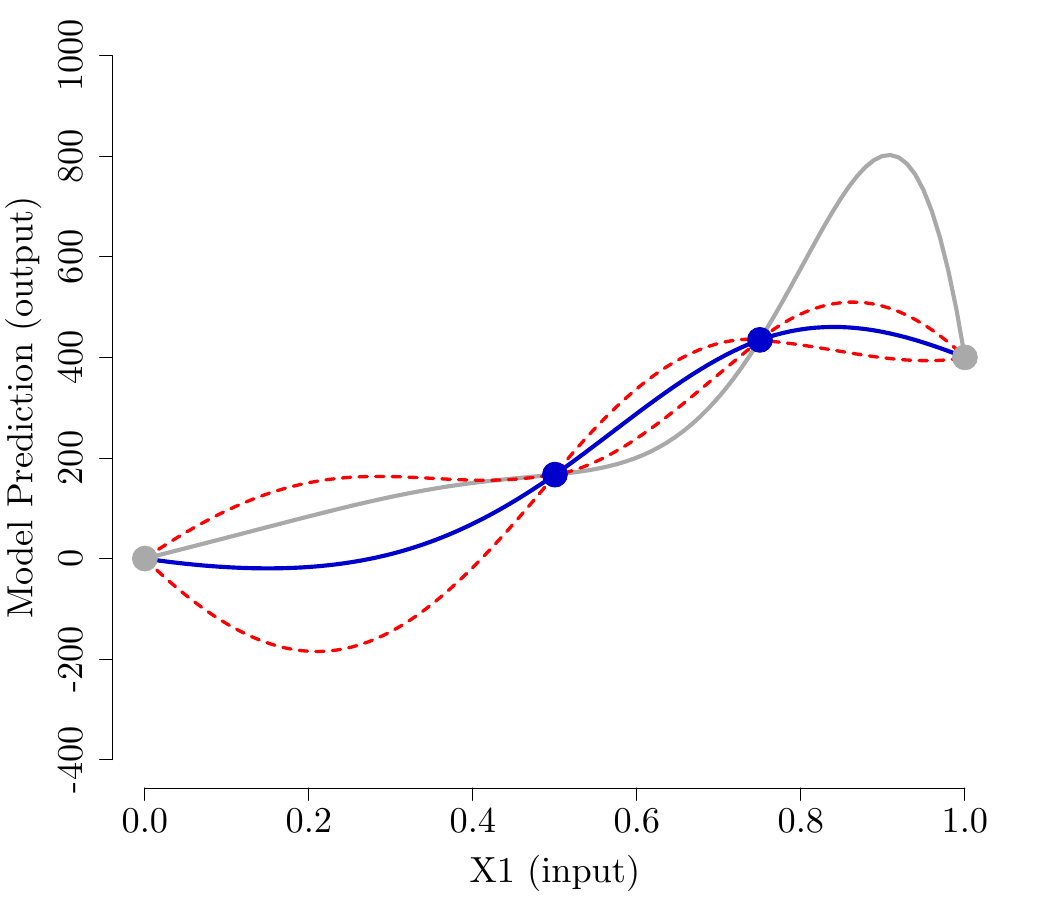} \subcaption{Remove $i$-th interior point (Step 1.1) and re-fit emulator $f^*_{(-i)}(\bm{x})$ (Step 1.2).}}
   \parbox[b][2in][t]{0.33\textwidth}{\includegraphics[width=.9\hsize]{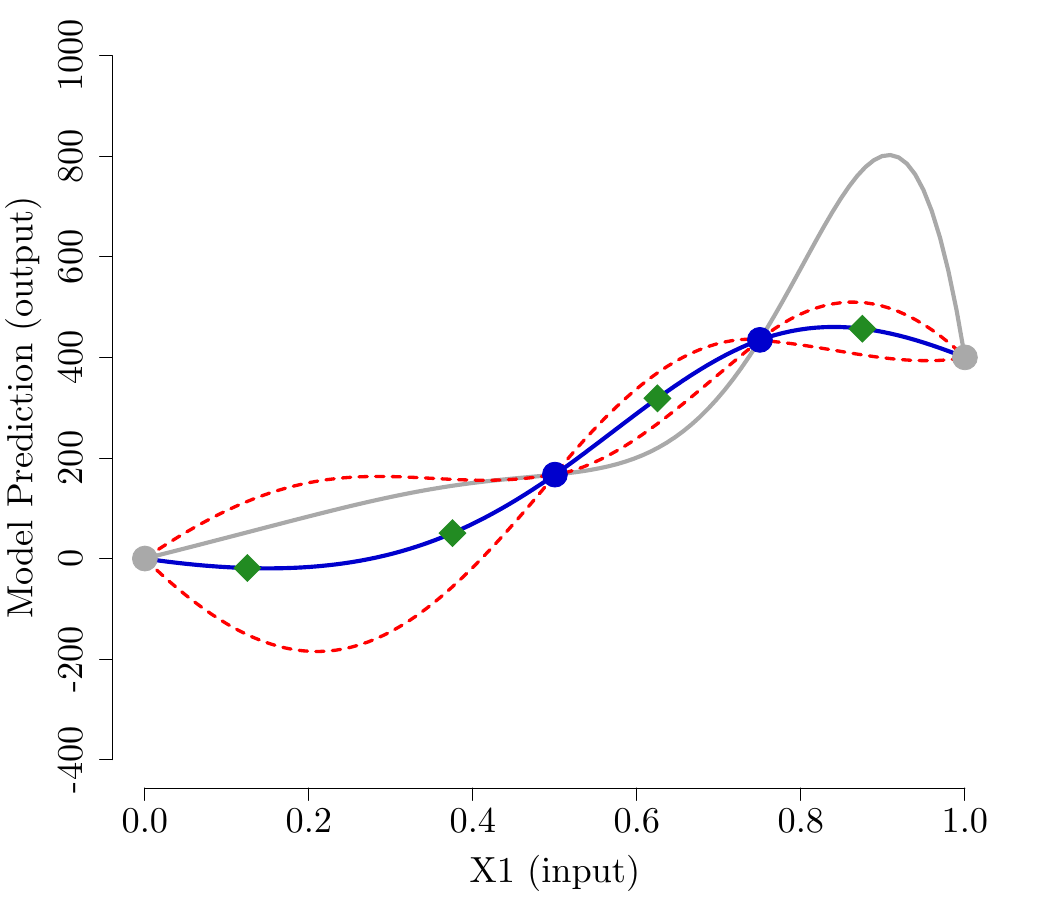} \subcaption{For each candidate input (Step 2), obtain predictions using re-fit emulators (Step 3.1).}}%
   \parbox[b][2in][t]{0.33\textwidth}{\includegraphics[width=.9\hsize]{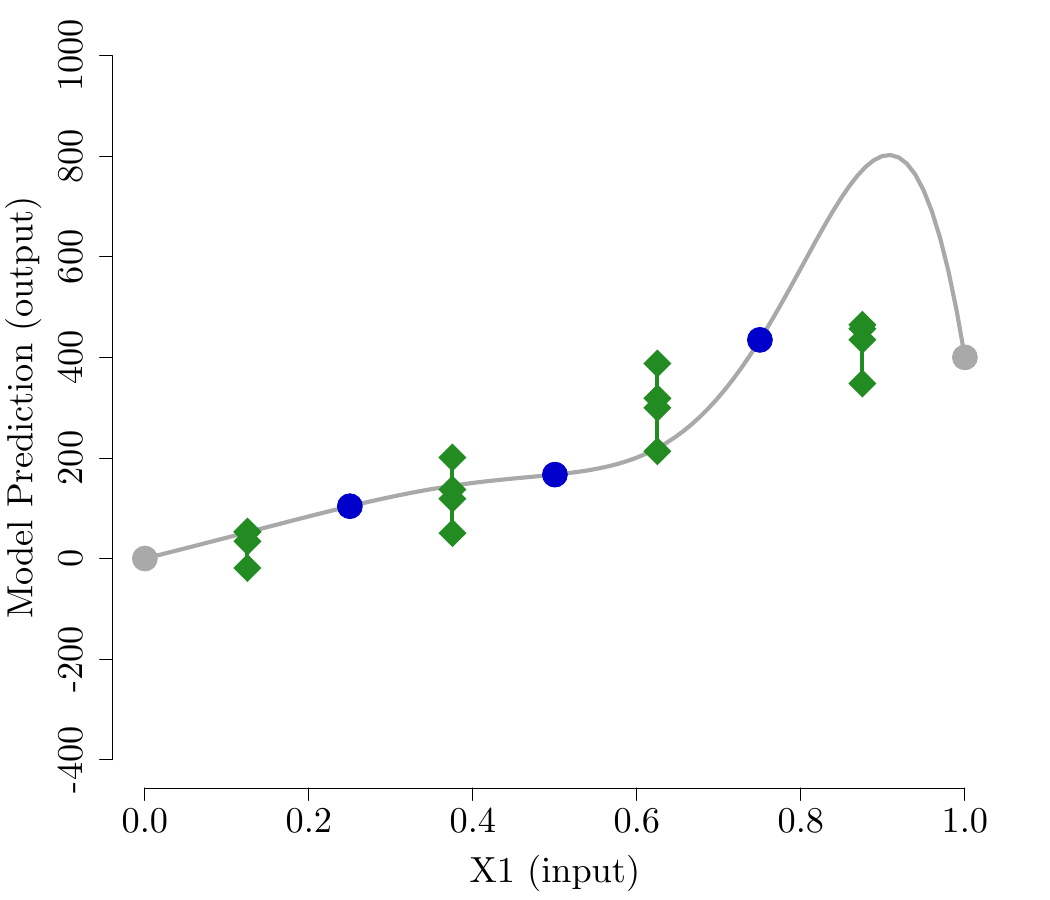} \subcaption{Obtain range of predictions at each candidate input over $f^*(\bm{x})$ and all $f^*_{(-i)}(\bm{x})$ (Step 3.2).}}\\[1ex]
   \parbox[b][2in][t]{0.33\textwidth}{\includegraphics[width=.9\hsize]{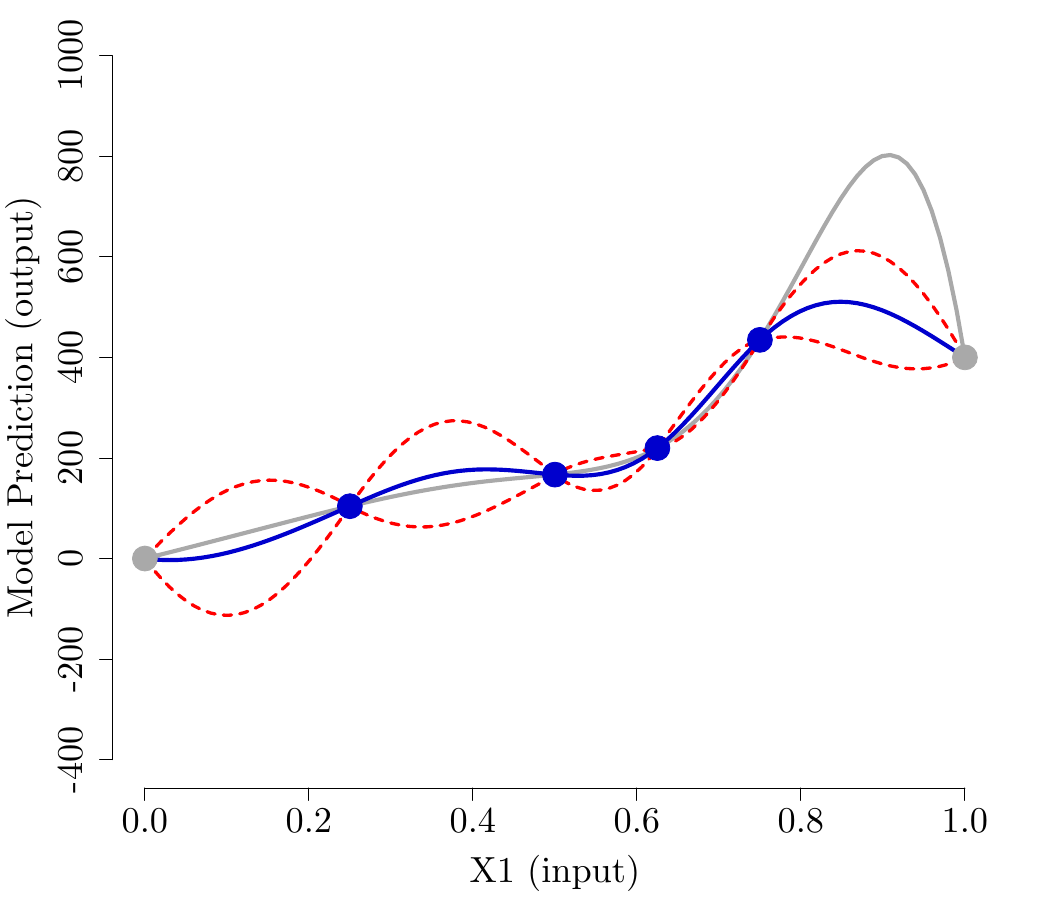} \subcaption{With the candidate input with the largest range in predictions (Step 4), fit a new emulator (Step 5).}}%
   \parbox[b][2in][t]{0.33\textwidth}{\includegraphics[width=.9\hsize]{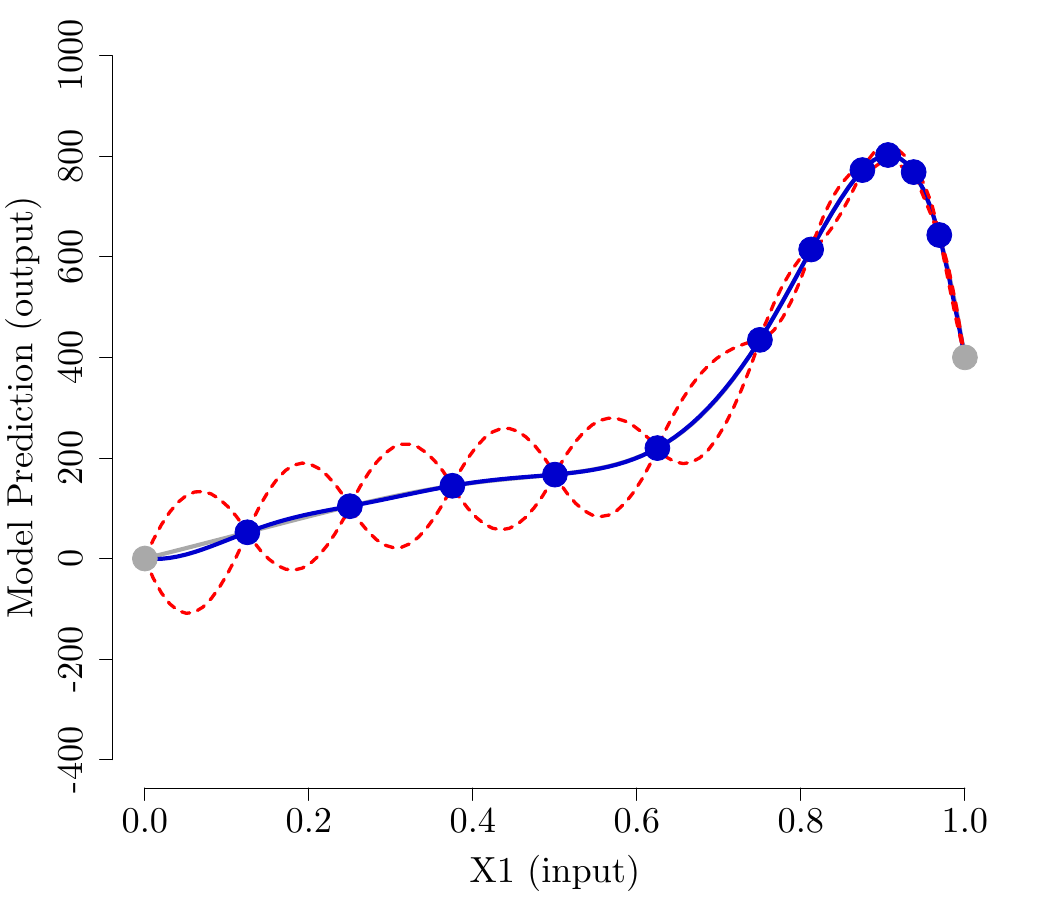} \subcaption{Repeat Steps 0-5 until range of predictions is below threshold.}}%
   \parbox[b][2in][t]{0.33\textwidth}{\includegraphics[width=.9\hsize]{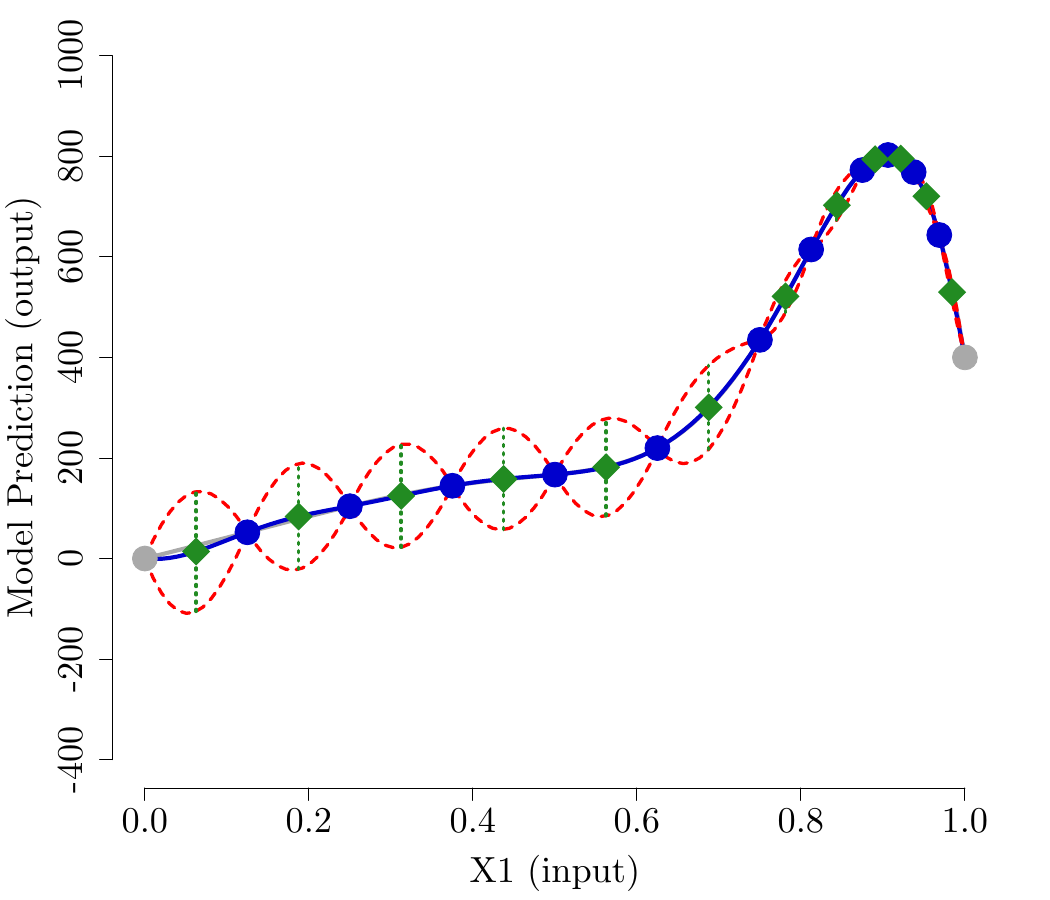} \subcaption{Estimate standard error of predictions (Step 5.1) and identify input with largest prediction error (Step 5.2).}}%
\caption[]{\textbf{1-dimensional example of ALGORITHM 1.} Blue points: interior design points; Grey points: design points corresponding to extreme vertices; Grey curve: simulator; Blue curve: emulator prediction. Red dashed curves: emulator's 95\% prediction interval bounds.
\label{Fig_AL}}
\end{figure}

\begin{featurebox}
\caption{\label{myBox1}DESIGN POINT ALGORITHM (ALGORITHM 1)}
Start with a seeding set $\mathcal{X}_{n_0}$ of $n_0$ input vectors. At least one vector in the set is in the interior of the polytope of $\mathcal{X}_{n_0}$. Call $\mathcal{D}_{n_0}$ the set of corresponding design points.  

\vspace{0.25cm}
For each iteration $t=0,1, \ldots \ $ until criteria are met:

\hspace{.25cm} $0.$ Fit an EMULATOR $\:f^*(\cdot)$ to all $n_0+t\:$ design points in $\mathcal{D}_{n_0+t}$

\hspace{.25cm} $1.$ For the $i$-th interior input vector $\bm{x}_i$:

\hspace{.75cm} $1.1.$ Remove the design point ($\bm{x}_i, f(\bm{x}_i))$  

\hspace{.75cm} $1.2.$ Refit EMULATOR $\:f_{(-i)}^*(\cdot)$ to the remaining design points in $\mathcal{D}_{n_0+t} \ \backslash \ \{ (\bm{x}_i, f(\bm{x}_i)) \}$ 

\hspace{.25cm} $2.$ Obtain a set $\mathcal{C}_t$ of candidate input vectors (see Section~\ref{sec:Algorithm2})

\hspace{.25cm} $3.$ For each candidate input vector $\bm{c}_t \in \mathcal{C}_t$: 

\hspace{.75cm} $3.1.$ Estimate predictions $f^*(\bm{c}_t)$, and $\:f_{(-i)}^*(\bm{c}_t)$ for all $\:f_{(-i)}^*(\cdot)$ in Step $1.2.$ Get the range $R(\bm{c}_t)$ of predictions. 


\hspace{.25cm} $4.$ Identify the candidate input vector $\bm{c}^*_{t} = \argmax \limits_{\bm{c}_{t} \in \mathcal{C}_{t}} \ R(\bm{c}_t)$
	   
\hspace{.25cm} $5.$ IF $R(\bm{c}^*_{t}) \ge {{T}_{resample}}\:$, where ${{T}_{resample}}$ is a predefined threshold:
  
\hspace{.75cm} - Add design point  $(\bm{c}^*_{t}, f(\bm{c}^*_{t}))$ to $\mathcal{D}_{n_0+t}$ and return to Step \textsf{0}.

\hspace{.6cm} ELSE: 

\hspace{.75cm} $5.1.$ Estimate the standard error of predictions $\:\hat{SE}(f^*(\bm{c}_t))$ at all $\bm{c}_t \in \mathcal{C}_t$ 

\hspace{.75cm} $5.2.$ Identify the candidate point $\bm{c}^{**}_{t}$  with $\argmax\limits_{\bm{c}_{t} \in \mathcal{C}_{t}}[\hat{SE}(f^*(\bm{c}_{t}))]$

\hspace{.75cm} $5.3.$ IF $\hat{SE}(f^*(\bm{c}^{**}_{t}))\ge {{T}_{SE}} \:$, where ${{T}_{SE}}$ is a predefined threshold:
  
\hspace{1.5cm} - Add design point $(\bm{c}^{**}_{t}, f(\bm{c}^{**}_{t}))$ to $\mathcal{D}_{n_0+t}$ and return to Step \textsf{0}.

\hspace{1.3cm} ELSE: END
\end{featurebox}
The proposed active learning (AL) algorithm (Box~\ref{myBox1}: ALGORITHM 1) adopts aspects of sequential designs from the current literature \cite{Kleijnen2008}. The algorithm seeks to select design points in the regions where (i) the simulator output changes quickly and (ii) the emulator has high predictive uncertainty. The first goal is pursued through a resampling procedure which successively removes existing design points from the complete set of design points, refits emulators using each of the resulting subsets of design points, and then identifies the ``candidate'' input vectors with the largest range in predictions obtained using this series of emulators. Candidate input vectors are input vectors that have not yet been evaluated with the simulator. With this resampling procedure, the regions where the simulator is ``fast-changing''  are likely to have larger ranges in predictions than in regions where the simulator is relatively flat \cite{Kleijnen2004}. For example, removing a design point from a region where small changes in a PSA threshold value produce large changes in life-years saved will likely change the prediction of a nearby PSA threshold value. If the resampling procedure does not identify any ``fast-changing'' regions (towards satisfying the first goal above), then the second goal is pursued by examining the regions with large variance in the emulator's prediction (this is, essentially, the ALM criterion). Choosing additional design points in the latter regions will reduce the emulator’s overall predictive uncertainty. Details of the algorithm’s process are described below, and Figure \ref{Fig_AL} illustrates the steps of the algorithm using a simplistic $1$-dimensional example.

To start, the algorithm requires an initial set $\mathcal{X}_{n_0}$ of $n_0$ input vectors, which includes the $n_*$ extreme vertices of the input space $\mathcal{X}$ and a non-empty set of $n_0 -n_*$ input vectors which are in the interior of the polytope of $\mathcal{X}$ (Figure~\ref{Fig_AL} (a)). The initial input vectors are paired with their output values evaluated in the simulator $f(\cdot)$ to obtain the set $\mathcal{D}_{n_0} = \{(\bm{x}, f(\bm{x})): \bm{x} \in \mathcal{X}_{n_0}\}$ of initial design points (Figure~\ref{Fig_AL} (b)). An emulator $f^*(\cdot)$ is then fit to $\mathcal{D}_{n_0}$ (Step $0$, Figure~\ref{Fig_AL} (c)). Next, for each interior input vector $\bm{x}_i$, the corresponding design point $(\bm{x}_i, f(\bm{x}_i))$ is removed from the complete set of design points (Step $1.1$) and a new emulator $f^*_{(-i)}(\cdot)$ is fit to the remaining set $\mathcal{D}_{n_0} \setminus \{(\bm{x}_i, f(\bm{x}_i)) \}$ of $n_0-1$ design points (Step $1.2$, Figure~\ref{Fig_AL} (d)). A set $\mathcal{C}$ of candidate input vectors with ``large'' prediction errors from the current emulator are then obtained (Step $2$). We describe a simple algorithm for selecting candidate points in Section~\ref{sec:Algorithm2}. For each candidate input vector $\bm{c} \in \mathcal{C}$, predictions are estimated using the emulator $f^*(\cdot)$ and each re-fit emulator $f^*_{(-i)}(\cdot)$ from Step 1.2. (Step $3.1$, Figure~\ref{Fig_AL} (e)); the range $R(\bm{c})$ of these predictions is obtained (Step $3.2$, Figure~\ref{Fig_AL} (f)). The ``winning'' candidate input vector $\bm{c^*} \in \mathcal{C}$ is the one with the largest range in predictions 
(Step $4$). If the range from this candidate input vector is above a pre-specified threshold (i.e., $R(\bm{c}^*)\ge {{T}_{resample}}$), the new design point $(\bm{c}^{*}, f(\bm{c}^{*}))$ is added to the set of existing design points and we return to Step $0$ (Step $5$, Figure ~\ref{Fig_AL} (g)). Step $0$ to Step $5$ are repeated until the range of predictions for $\bm{c^*}$ is below the threshold ${{T}_{resample}}$ (Figure ~\ref{Fig_AL} (h)). Then, the standard error of the predictions $\hat{SE}(f^*(\bm{c}))$ with emulator $f^*(\cdot)$ is estimated for each candidate input vector (Step $5.1$, Figure ~\ref{Fig_AL} (i)), and the candidate input vector $\bm{c}^{**} \in \mathcal{C}$ with the largest prediction error 
is identified (Step $5.2$). If the prediction error for this candidate input vector is above a pre-specified threshold (i.e., $\hat{SE}(f^*(\bm{c}^{**}))\ge {{T}_{SE}}$),  then the new design point $(\bm{c}^{**}, f(\bm{c}^{**}))$) is added to the set of existing design points and we return to Step $0$ (Step $5.3$). The process repeats until both criteria are satisfied, possibly for several (say 5) consecutive new design points.

In terms of the attributes of active learning algorithms in Section~\ref{sec:emulator-based_designs}, observe that: (1) The active learning criteria involve selecting the point that minimizes measures of variability of the GP mean and the predictive variance of the GP (akin to the ALM criterion). To reduce the dependency of the algorithm on emulators' ``overfiting'' bias, we make use of a resampling procedure that estimates these measures over a set of jack-knifed emulators. (2) We optimize the active learning criteria over the input space using a partitioning and local search algorithm (Section~\ref{sec:Algorithm2}). (3) When two design points are too close to each other, the GP estimation algorithms may fail to converge for numerical reasons. To avoid this difficulty, we fit GPs with a nugget (regularizer) term, estimated using the iterative scheme in~\cite{ranjan2011}. 

\subsection{Algorithm for candidate points}\label{sec:Algorithm2}

\begin{figure}
\captionsetup[subfigure]{margin={0cm,0cm}}
   \parbox[b][2.5in][b]{0.6\textwidth}{\includegraphics[width=\hsize]{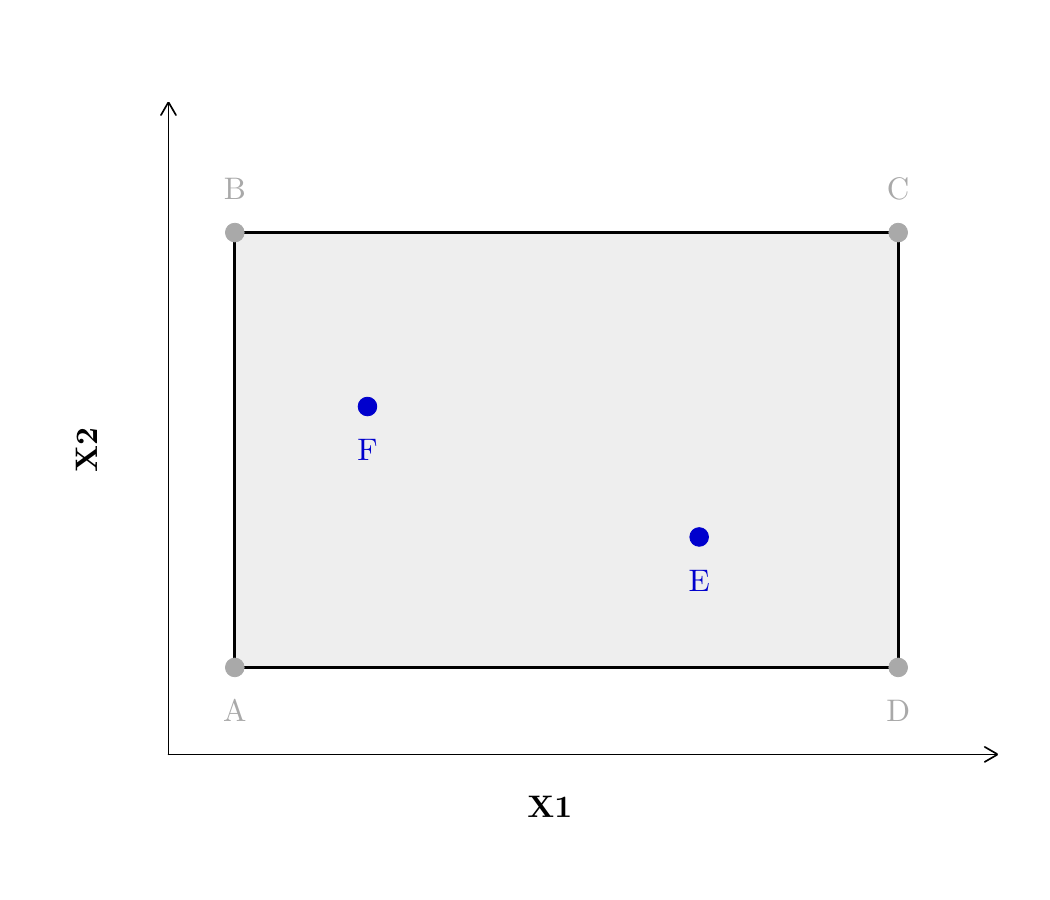}}%
   \parbox[b][2.5in][t]{0.4\textwidth}{\subcaption{\textbf{Initial set-up: Input vectors.} \\
   \vspace{.2cm}
   The polytope of $\mathcal{X}_6$ is the grey shaded rectangle, which includes the extreme vertices A, B, C, D (grey points). E, F are interior vectors (blue points).}}\\[3ex]
   \parbox[b][2.5in][b]{0.6\textwidth}{\includegraphics[width=\hsize]{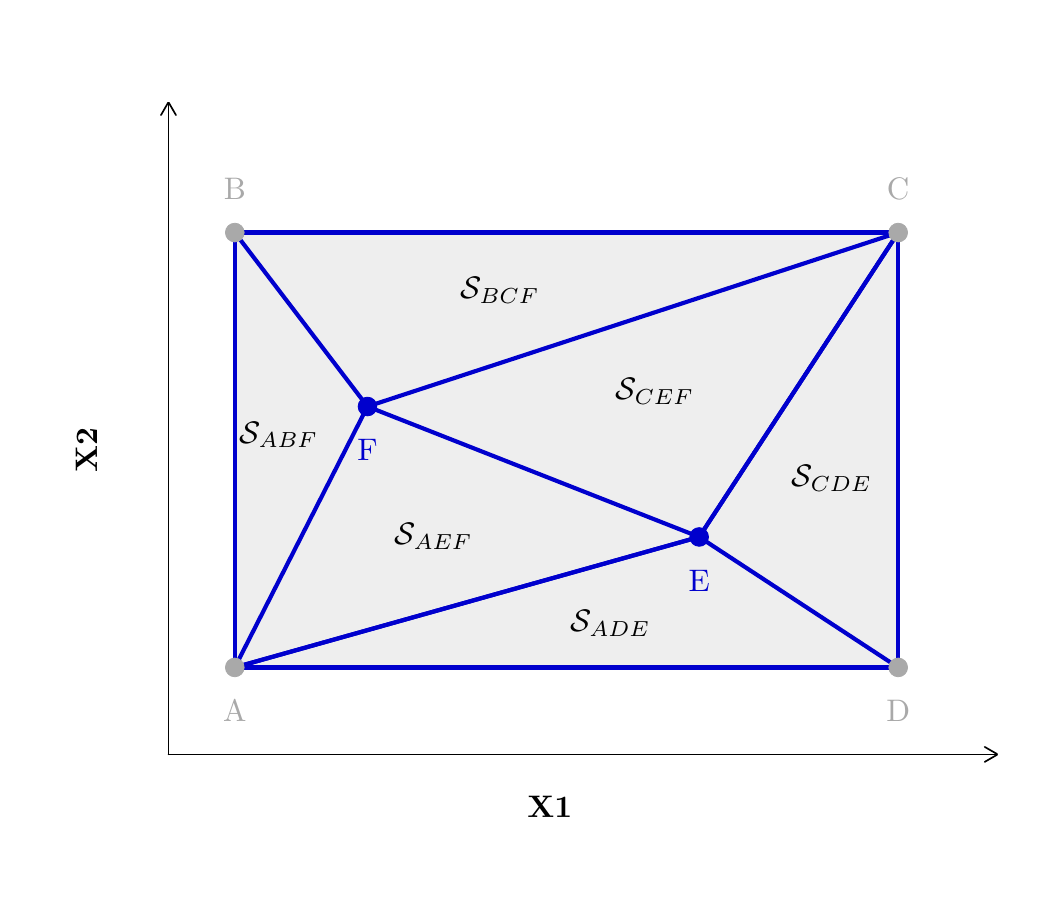}}%
   \parbox[b][2.5in][t]{0.4\textwidth}{\subcaption{\textbf{Step 1.} \\
   \vspace{.2cm}
   The triangulation gives the 6 simplexes $T(\mathcal{X}_6)=\{\mathcal{S}_{ADE}, \mathcal{S}_{CDE},$  $\mathcal{S}_{CEF},\mathcal{S}_{BCF},\mathcal{S}_{AEF},\mathcal{S}_{ABF} \}$ defined by the blue line segments.}}\\[3ex]
   \parbox[b][2.5in][b]{0.6\textwidth}{\includegraphics[width=\hsize]{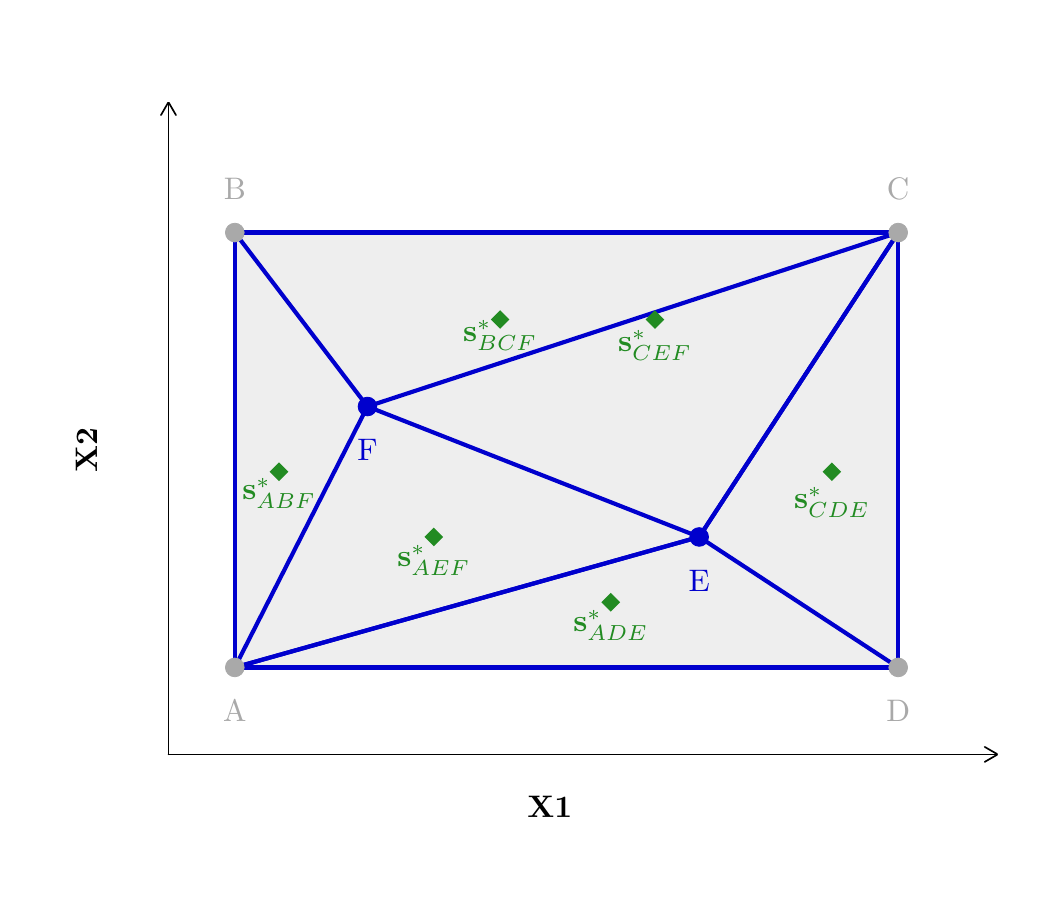}}%
   \parbox[b][2.5in][t]{0.4\textwidth}{\subcaption{\textbf{Step 1.1.}\\
   \vspace{.2cm}
   Starting from the centroid of each of the 6 simplexes, identify the vectors $\{\bm{s}^*_{ADE}, \bm{s}^*_{CDE},\bm{s}^*_{CEF},$ $\bm{s}^*_{BCF}, \bm{s}^*_{AEF},\bm{s}^*_{ABF}\}$ (green diamonds) that are local maxima of the standard error of the predictions over $\mathcal{X}_6$.}}
\caption{\textbf{Illustrative example of ALGORITHM 2.}
\label{fig_AL2}}
\end{figure}

The identification of the set $\mathcal{C}$ of candidate input vectors for ALGORITHM $1$ is outlined in ALGORITHM 2 (Box~\ref{myBox2}). Figure~\ref{fig_AL2} provides an illustration in a $2$-dimensional example. To start, the algorithm requires a set $\mathcal{D}_{n}$ of design points (Figure~\ref{fig_AL2} (a)), with $\mathcal{X}_{n}$ denoting the corresponding set of input vectors, and an emulator $f^*(\cdot)$ fit with $\mathcal{D}_{n}$. 

First, a triangulation of $\mathcal{X}_{n}$ is obtained (Step 1, Figure~\ref{fig_AL2} (b)). 
The triangulation $T(\mathcal{X}_n) = \cup \mathcal{S}_j$ partitions $\mathcal{X}_{n}$ in $k-$dimensional simplexes $\mathcal{S}_j$ using all $\bm{x} \in \mathcal{X}_{n}$ as vertices. The partitioning is such that $\mathcal{S}_j \cap \mathcal{S}_l$ for $j\neq l$ is either the empty set or a lower-dimensional simplex (a shared extreme point, line, or facet). 

Any triangulation would work. For relatively small numbers of points (in the few hundreds), and few dimensions (say, less than 10), it is practical to use a Delaunay triangulation, for which many routines exist~\cite{Delaunay1934}. Within each returned simplex, we identify the vector $\bm{s}_j^*$ that maximizes the standard error of the predictions $\hat{SE}(f^*(\bm{c}))$ with emulator $f^*(\cdot)$, for all $\bm{s}_j \in \mathcal{S}_j$, i.e., the ALM criterion (Step $1.1$ in ALGORITHM 1, Figure~\ref{fig_AL2} (c)). The set $\mathcal{C}$ of unique $\bm{s}_j^*$ are selected as the candidate input vectors to be exploited in ALGORITHM $1$. 

\begin{featurebox}
\caption{CANDIDATE DESIGN POINT ALGORITHM (ALGORITHM 2)}

For a given set $\mathcal{D}_n$ of design vectors, let $\mathcal{X}_n$ be the set of the corresponding input vectors, and $f^*(\cdot)$ an emulator fit with $\mathcal{D}_n$.   
 
\vspace{0.25cm}
$1.$ Obtain a triangulation $T(\mathcal{X}_n)$. For each simplex $\mathcal{S}_j$ in $T(\mathcal{X}_n)$: 

\hspace{.25cm} $1.1$ Find the input vector $\mathbf{s}_j^*$ that maximizes the standard error of the prediction $\:\hat{SE}(f^*(\mathbf{s}_j))$ for all $\mathbf{s}_j \in  \mathcal{S}_j$.   

$2.$ Return the candidate input vectors $\mathcal{C} = \textrm{unique} \big ( \{ \mathbf{s}^*_j \textrm{for all } j \} \big )$ from the input vectors identified in Step 1. 
\label{myBox2}
\end{featurebox}

\subsection{Implementation}\label{sec:implementation}

Our active learning algorithm is implemented in {\fontfamily{qcr}\selectfont R}~\cite{R2019} and uses a modified version of the {\fontfamily{qcr}\selectfont GPfit} package for fitting GP emulators~\cite{Macdonald2015,ranjan2011}. We fit stationary GPs with a nugget term, whose magnitude is specified with the iterative Tikhonov regularization algorithm in~\cite{ranjan2011}. When refitting emulators during iterations of the active learning algorithm, we used the initial values of previously fit emulators as initial values.  We generated LHS and MaxPro space-filling designs with the {\fontfamily{qcr}\selectfont lhs}~\cite{lhs} and {\fontfamily{qcr}\selectfont MaxPro}~\cite{joseph2015} packages in {\fontfamily{qcr}\selectfont R}. We used the {\fontfamily{qcr}\selectfont tgp} package to implement the Bayesian treed GPs~\cite{gramacy2007,gramacy2010}. We used the {\fontfamily{qcr}\selectfont geometry} package~\cite{Barber2012} to obtain the triangulation of $\mathcal{X}_n$ that is required for the algorithm in Box~\ref{myBox2} that identifies the candidate points at each iteration of the active learning algorithm, and the {\fontfamily{qcr}\selectfont lhs} package to generate LHS designs~\cite{lhs}. 
Code to run the benchmark examples is provided online.\footnote{See \url{https://github.com/ttrikalin/AL_for_emulators_paper1}}

\section{Experiments}\label{sec:examples}

\subsection{Experimental setup}

We compared the performance of our algorithm with state-of-the-art emulator-free and emulator-based (active learning) algorithms using seven benchmark functions and the PSAPC model.  
We recorded the evolution of the square root mean square error (RMSE) and the maximum deviation (MAX) between simulators and emulators fit with $n$ design points ranging from $n_0$ to a maximum of $B=100$. 
The emulator-free designs (MaxPro for the benchmark functions, and both MaxPro and random LHS for PSAPC) would use $n=B$ design points, but we show curves for $n_0 \leq n \leq B$. The active learning algorithms (bayesian treed GPs with ALM and ALC stopping criteria, and ALGORITHM 1) start with a seeding design of size $n_0$ and augment it until convergence or until the budget is exhausted. 

Table~\ref{tab:experimental_setup} summarizes the experimental setup. The benchmark function analyses involved minimal exploration, as they are not models of phenomena: We generated curves for performance metrics starting with a single set $n_0=3k$ design points, did not run random LHS designs and ran one replicate for ALGORITHM 1, which is deterministic, and 10 replicates for each treed GP-based active learning algorithm, because their estimation involves Markov Chain Monte Carlo computations.

For PSAPC, we did more extensive analyses: We generated curves for performance metrics for starting designs with sizes of $n_0=3k$ or $10k$, to see if starting from denser designs is advantageous. For each choice for $n_0$, we run 100 random LHS algorithms (10 starting sets with randomly chosen internal points, each repeated 10 times). We run one replicate of ALGORITHM 1 starting from each of 11 seeding designs (the same 10 as with the LHS and one where the external design vectors were augmented with a MaxPro design), and 10 replicates for each treed GP-based algorithm for the same 11 starting designs as with ALGORITHM 1. 

\begin{table}[t!]
\rowcolors{1}{}{}
\caption{\textbf{Experimental setup}}
\label{tab:experimental_setup}
\centering
\resizebox{\textwidth}{!}{%
\begin{tabular}{lcc|ccc}
\textbf{Attribute (per type of experiment, if different)} & \multicolumn{2}{c}{\textbf{Emulator-free designs}} & \multicolumn{3}{c}{\textbf{Emulator-based design algorithms}} \\
 & \textbf{MaxPro} & \textbf{Random LHS} & \textbf{treed GPs (ALC)} & \textbf{treed GPs (ALM)} & \textbf{ALGORITHM 1} \\ \hline
Emulator mean function  ($m(\cdot)$) & constant & constant & constant & constant & constant \\ \hline
Nugget used & yes & yes & yes & yes & yes \\ \hline
Number of input variables ($k$) &  &  &  &  &  \\
$ \  \  \ $Benchmark functions & 1, 2, or 3 & not done & 1, 2, or 3 & 1, 2, or 3 & 1, 2, or 3 \\
$ \  \  \ $PSAPC & 1, 2, 3, 4 & 1, 2, 3, 4 & 1, 2, 3, 4 & 1, 2, 3, 4 & 1, 2, 3, 4 \\ \hline
Number of initial design points ($n_0$) &  &  &  &  &  \\
$ \  \  \ $Benchmark functions & $3k$ & not done & $3k$ & $3k$ & $3k$ \\
$ \  \  \ $PSAPC & $3k, 10k$ & $3k, 10k$ & $3k, 10k$ & $3k, 10k$ & $3k, 10k$ \\ \hline
Number of initial seeding sets of design points &  &  &  &  &  \\
$ \  \  \ $Benchmark functions & 1 & not done & 1 & 1 & 1 \\
$ \  \  \ $PSAPC & 1 & 10 & \begin{tabular}[t]{@{}l@{}}11*\end{tabular} & \begin{tabular}[t]{@{}l@{}}11* \end{tabular} & \begin{tabular}[t]{@{}l@{}}11*\end{tabular} \\ \hline
Number of random runs per initial set of design points** &  &  &  &  &  \\
$ \  \  \ $Benchmark functions & 1 & not done & 10 & 10 & 1 \\
$ \  \  \ $PSAPC & 1 & 10 & 10 & 10 & 1 \\ \hline
%
Maximum design point budget ($B$) & 100 & 100$^\text{\cmark}$ & 100 & 100 & 100 \\ \hline
Convergence threshold (for sequential algorithms) &  &  &  &  &  \\
$ \  \  \ $Benchmark functions & NA & NA & $\Delta\text{ALC}<10^{-4}, 10^{-5}$ & \begin{tabular}[t]{@{}l@{}} ALM$<7.5\%$ of\\output range$^\text{\dmark}$ \end{tabular} & \begin{tabular}[t]{@{}l@{}}$T_{resample}=T_{SE} <$7.5\%\\of output range$^\text{\dmark}$\end{tabular} \\
$ \  \  \ $PSAPC & NA & NA & $\Delta\text{ALC}<10^{-5}$ & ALM$<0.5$ QAD & \begin{tabular}[t]{@{}l@{}}$T_{resample}=$0.2 QAD\\$T_{SE}=$ 0.5 QAD\end{tabular} \\ \hline
\begin{tabular}[t]{@{}l@{}}Number of consecutive iterations \\ meeting threshold criteria (for sequential algorithms)\end{tabular} & NA & NA & 1,2,3,4,5 & 1,2,3,4,5 & 1, 2, 3,4, 5  \\ \hline
\end{tabular}%
}
\begin{tablenotes}
\begin{footnotesize}
\item $^*$ The same 10 starting seeds as the 10 random LHS, plus a starting set obtained with a MaxPro design.
\item $^{**}$ LHS requires random sampling and Bayesian treed GPs require Markov Chain Monte Carlo computation; they are repeated 10 times for each set of starting points. 
\item $^\text{\cmark}$ In the PSAPC experiments, the LHS simulation selects points in batches of 5; it stopped selecting design points when 100 points had been selected. For $n_0=3k$, the final number of design points with LHS was 103 ($k=1$), 101 ($k=2$), 104 ($k=3$), and 102 ($k=4$).
\item $^\text{\dmark}$ The threshold was 7.5\% of the output range for functions with 1 input and 3 to 5\% of output range for functions with 2 and 3 inputs.
\item ALC|ALM: Active learning with the Cohn~\cite{cohn1994}|MacKay~\cite{mackay1992} criterion; LHS: Latin hypercube sampling; NA: Not applicable; QAD: Quality adjusted days
\end{footnotesize}
\end{tablenotes}
\end{table}

\subsection{Benchmark functions}\label{sec:benchmarks}
\begin{figure}[bt!]
 \centering
\includegraphics[width=\textwidth]{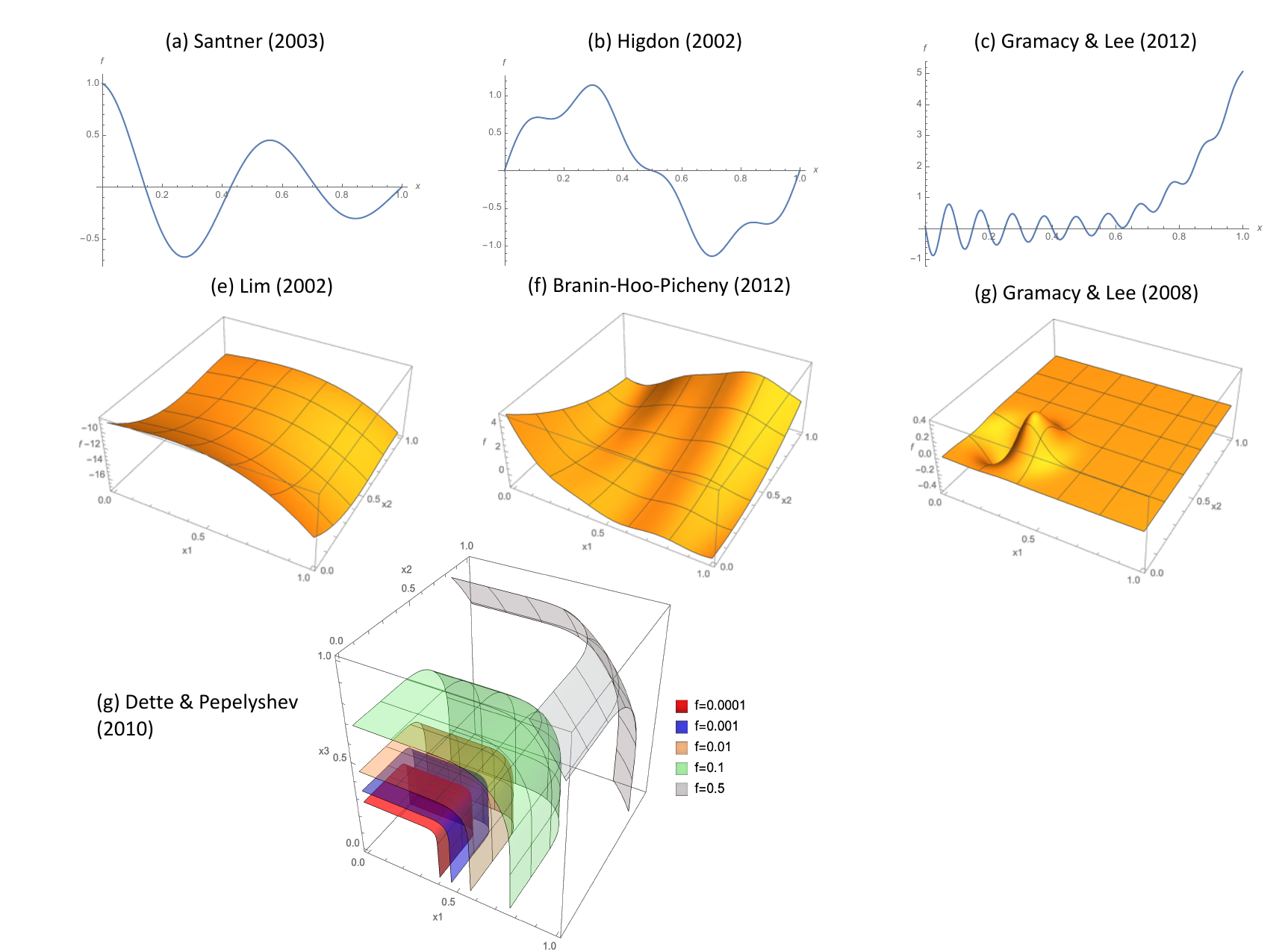}
\caption[]{Benchmark functions used in the experiments. Functions (a) to (c) have one input in $[0,1]$, and (d) to (f) have two inputs in $[0, 1]^2$. For function (g), which has three inputs in $(0,1]^3$, we plot the iso-contours for values of the function equal to $10^{-4}, 10^{-3}, 10^{-2}, 10^{-1},$ and $0.5$. The seven functions vary in their smoothness over areas of the input domain in different ways. The functions are from the following publications: (a)~\cite{santner2003}; (b)~\cite{higdon2002}, (c)~\cite{gramacy2012}, (d)~\cite{lim2002}, (e)~\cite{picheny2013}, (f)~\cite{gramacy2008b,gramacy2008}, (g)~\cite{dette2010}.  
\label{fig:benchmark_functions}}
\end{figure}

We use the seven benchmark functions in Figure~\ref{fig:benchmark_functions}, which pose various challenges for emulation. Their specifications and extreme points over the input domain are shown in Appendix~\ref{app:benchmark_specs}. All functions are highly non-linear and some exhibit cyclical behavior with varying amplitude. In Figure~\ref{fig:benchmark_functions}, we arrange functions in terms of increasing number of input dimensions and non-smoothness. Function (a) is a relatively low frequency wave with exponentially decreasing amplitude. Function (b) is the sum of a larger amplitude lower frequency wave and a smaller amplitude higher frequency wave and has many inflection points. Function (c) is the sum of a 4th degree polynomial and a relatively high-frequency wave with exponentially decaying amplitude. Its smoothness varies more over the input domain compared to (a) and (b). Of the functions with two dimensional inputs, function (f) is remarkable in that its smoothness varies much more in the $[0, 0.5]^2$ part of its input domain than in the remaining part of the unit square. The three dimensional function decreases exponentially at different rates in the three dimensions. 

A primary goal was to emulate the benchmark functions with a square root mean square error (RMSE) or maximum deviation (MAX) smaller than 5\% of the maximum range for functions with one or more than one input dimension, respectively. For active learning algorithms, a secondary goal was to jointly minimize performance error and the number of design points at convergence.

\subsection{The PSAPC model}\label{sec:PSAPC_model}
The PSAPC microsimulation model, described in more detail in Appendix~\ref{app:psapc_model}, accounts for the relationship between PSA levels, prostate cancer disease progression, and clinical detection \cite{CISNETpc,Gulati2010}. The model, its estimation approach, its calibration, and its comparison with other prostate cancer models have been described in detail elsewhere \cite{Gulati2010, Gulati2013, Gulati2014, psapc2009}. Here, we treat the PSAPC model as a ``black box''.  The PSAPC model estimates the means of several clinical and procedural outcomes by forward Monte Carlo in simulated cohorts of men~\cite{Gulati2013}. To keep the Monte Carlo error negligible, we simulated cohorts of 100 million men aged 40 in the year 2000. Each simulation took approximately 15 minutes per processor thread on a cluster with a plurality of 1.6 GHz multicore processor nodes.

With the PSAPC model, we estimated the expected life-days saved with PSA-based screening versus no screening for policy-relevant screening strategies that used age-specific PSA thresholds. We considered all annual PSA screening strategies that used PSA positivity thresholds between 0.5 and 6.0 ng/mL and used four age-specific PSA positivity thresholds ($k=4$): one for men aged 40-44 years ($PSA_{40-44}$), a second for men aged 45-49 years ($PSA_{45-49}$), a third for men aged 50-54 ($PSA_{50-54}$), and a fourth for men aged 55-74 ($PSA_{55-74}$). Because PSA levels increase with increasing age \cite{Oesterling1993}, we assumed ``policy-relevant'' strategies consist of those with age-specific PSA positivity thresholds being constant in each age-group, where the PSA positivity threshold value for a given age group was at least as high as the value in the preceding age group. Thus, the input space is a simplex.  In sensitivity analyses, we also considered each of $k=1,2,3$ age-specific PSA positivity thresholds.

\subsection{Performance metrics}
For each experiment, we assessed each emulator's ``closeness'' to the respective emulator by estimating two metrics over a fine grid of the input space: 
(i) The square-root mean squared error RMSE$=\sqrt{\frac{1}{M}\sum_{m=1}^{M} (f^*(\bm{x}_m) - f(\bm{x}_m))^{2}}$, where $m=1,\dots,M$ indexes the grid points in the inputs. RMSE averages discrepancies over the whole input domain.  (ii) The maximum difference between the emulator prediction and the PSAPC simulator output, MAX$=\max_m|f^*(\bm{x}_m) - f(\bm{x}_m)|$.
For both metrics, a lower value indicates better performance (i.e., closer to the benchmark simulators or the PSAPC simulator). 
For experiments with benchmark functions, we scaled RMSE and MAX (sRMSE and sMAX, respectively) so that their maximum value is 1. All benchmark function inputs are in the unit interval, square or cube. We used $201$ grid points for functions with one input, and $10,000$ for functions with 2 and 3 inputs. For experiments with the PSAPC simulator, PSA positivity threshold values ranged from 0.5 to 6 ng/mL, with the PSA positivity threshold value for a given age group at least as high as the value in the preceding age group, as described above. For $k=4$, the number of evaluations was $M=11,616$.%
\footnote{The number of evaluations is as follows. For $k=1$, there were 101 evaluations so that each $PSA_{40-74}$ interval had length 0.01, yielding $M=101$. For $k=2$, a fine grid with intervals of 0.02 was obtained, then restricted to the subset where the values for $PSA_{45-74}\geq PSA_{40-44}$. The 101 values from $k=1$ were also included for a total of $M=1,376$ evaluations. For $k$, a fine grid with intervals of 0.04 was obtained, then restricted to the subset where the values for $PSA_{50-74}\geq PSA_{45-49}\geq PSA_{40-44}$. The 1,376 values from $k=2$ were also included for a total of $4,301$ evaluations. For $k=4$, a fine grid with intervals of 0.05 was obtained, then restricted to the subset where the values for $PSA_{55-74}\geq PSA_{50-54} \geq PSA_{45-49} \geq PSA_{40-44}$. The 4,301 values from $k=3$ were also included for a total of $11,616$ grid points.}

\section{Results}\label{sec:results}
\subsection{Results with benchmark functions}

\begin{figure}[t!]
\captionsetup[subfigure]{margin={0cm,0cm}}
   \parbox[c][1.3in][t]{0.49\textwidth}{\includegraphics[width=.95\hsize, trim=0 0 0 2cm, clip]{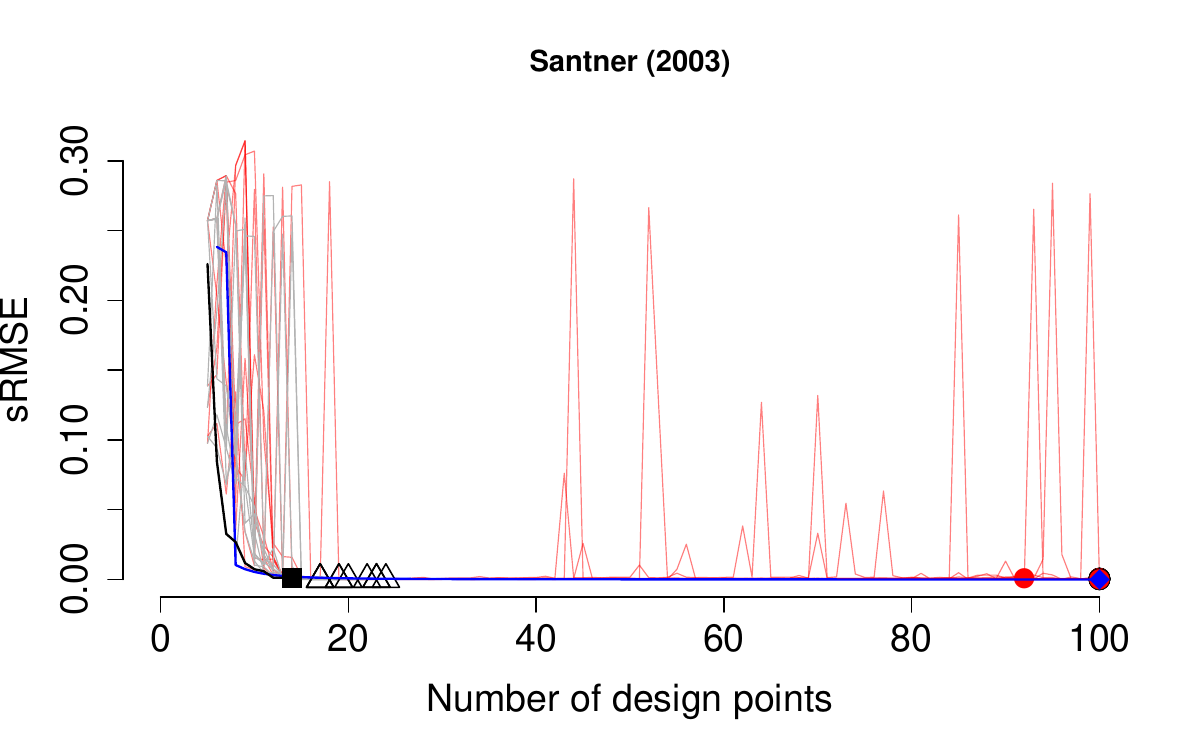}}%
   \parbox[c][1.3in][t]{0.49\textwidth}{\includegraphics[width=.95\hsize, trim=0 0 0 2cm, clip]{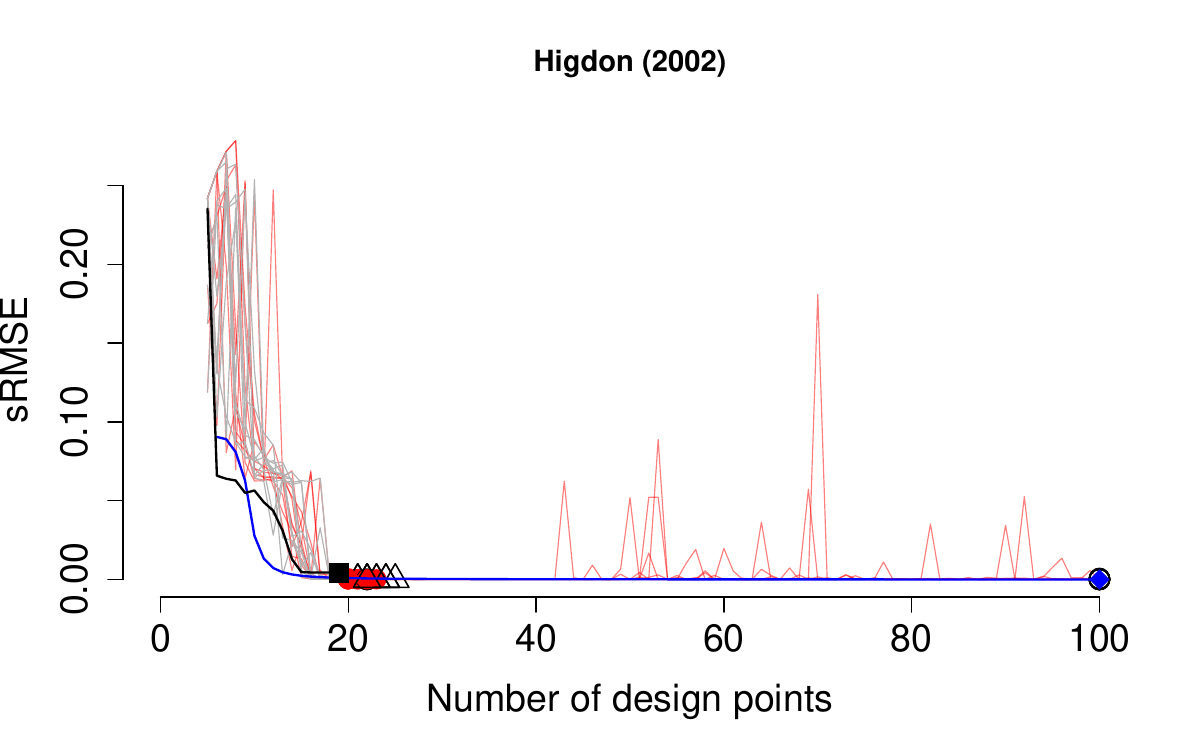}}\\[4ex]
   \parbox[t][.1in][c]{0.49\textwidth}{\subcaption{Santner (2003)}}
   \parbox[t][.1in][c]{0.49\textwidth}{\subcaption{Higdon (2002)}}\\[4ex]
   \parbox[c][1.3in][t]{0.49\textwidth}{\includegraphics[width=.95\hsize, trim=0 0 0 1.8cm, clip]{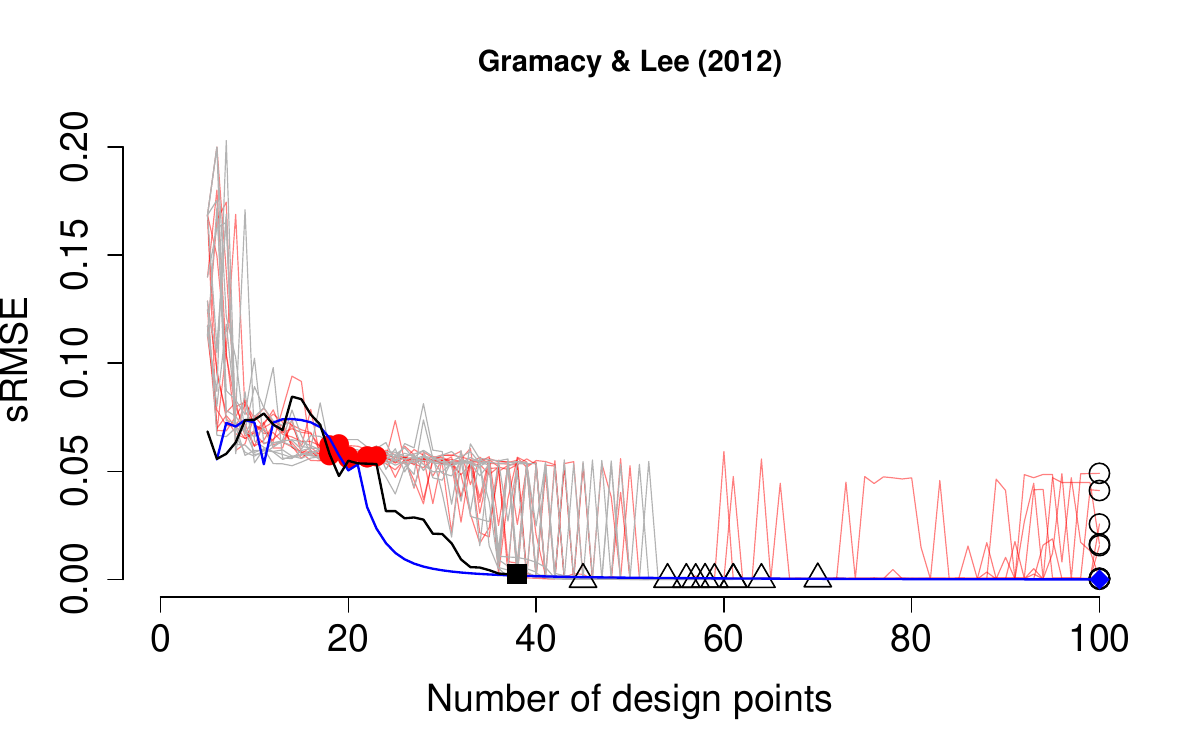}}%
   \parbox[c][1.3in][t]{0.49\textwidth}{\includegraphics[width=.95\hsize, trim=0 0 0 1.8cm, clip]{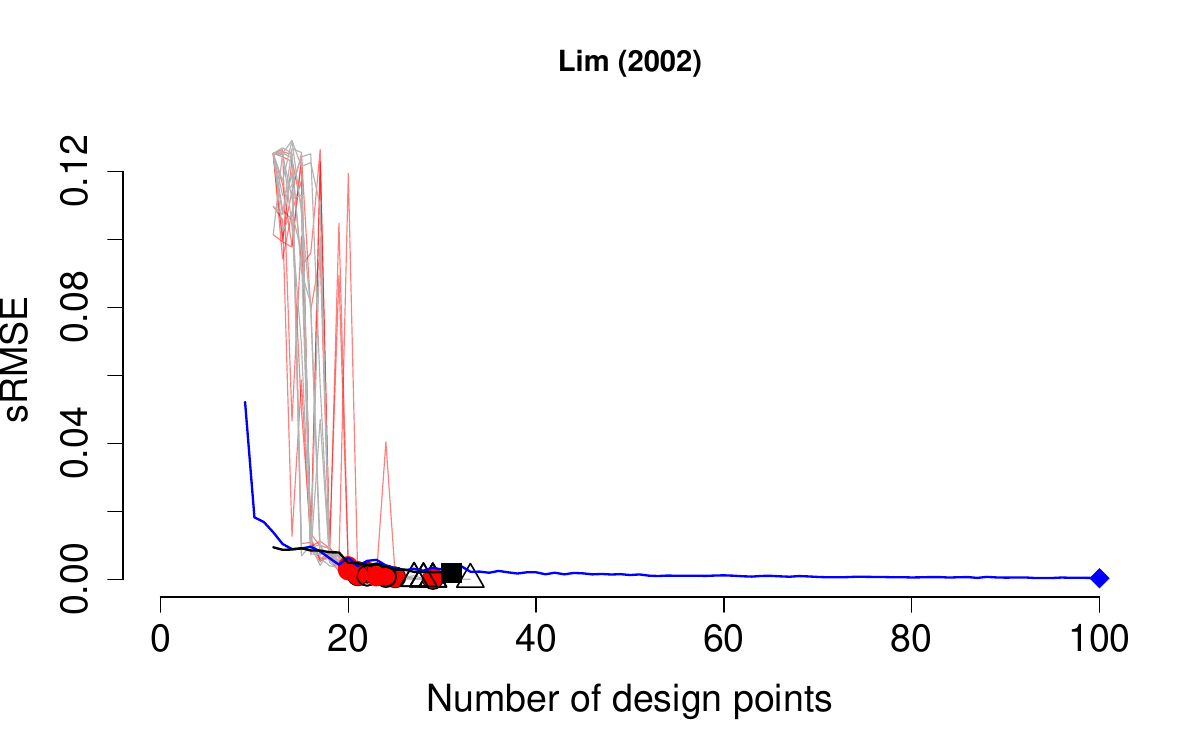}}\\[4ex]
   \parbox[t][.1in][c]{0.49\textwidth}{\subcaption{Gramacy \& Lee (2012)}}
   \parbox[t][.1in][c]{0.49\textwidth}{\subcaption{Lim (2002)}}
   \parbox[c][1.3in][t]{0.49\textwidth}{\includegraphics[width=.95\hsize, trim=0 0 0 2cm, clip]{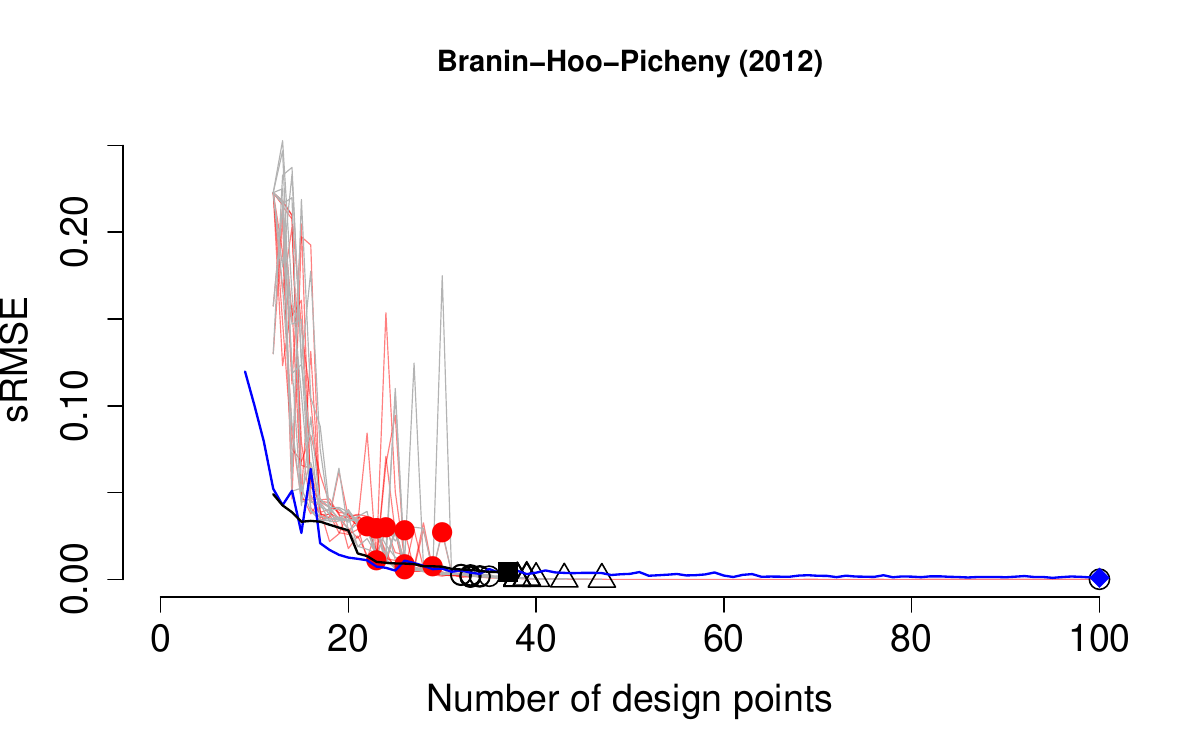}}%
   \parbox[c][1.3in][t]{0.49\textwidth}{\includegraphics[width=.95\hsize, trim=0 0 0 2cm, clip]{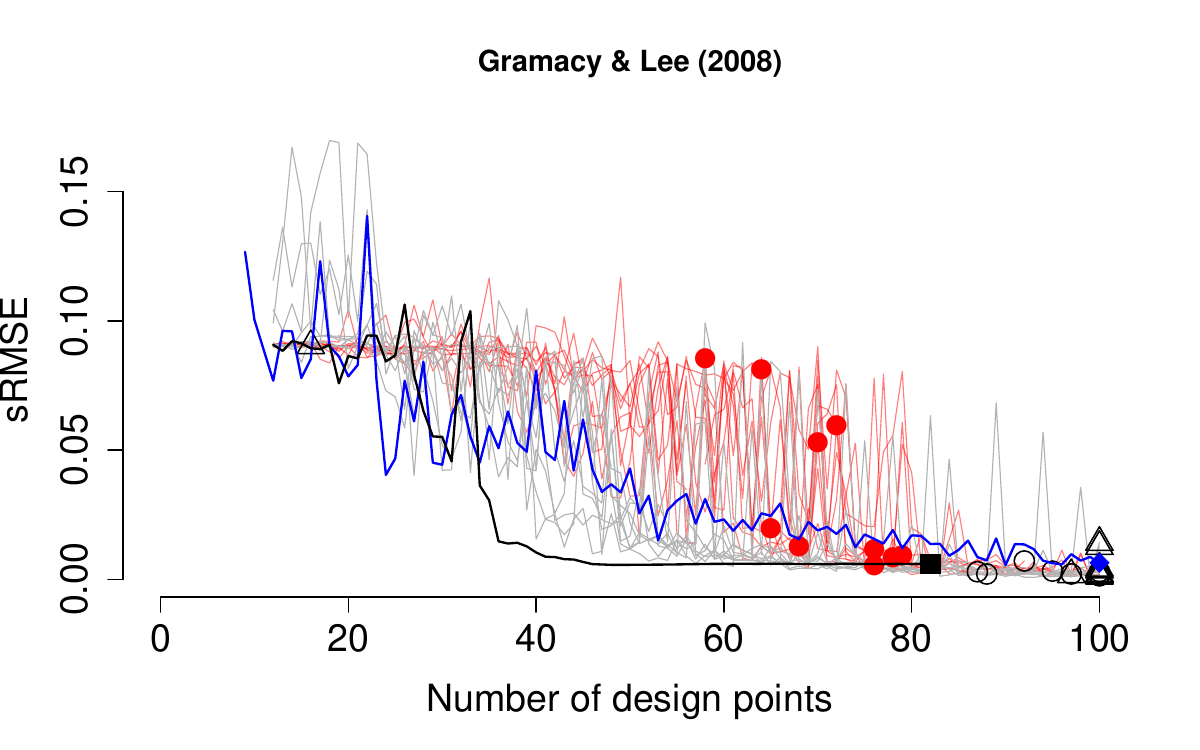}}\\[4ex]
   \parbox[t][.1in][c]{0.49\textwidth}{\subcaption{Branin-Hoo-Picheny (2012)}}
   \parbox[t][.1in][c]{0.49\textwidth}{\subcaption{Gramacy \& Lee (2008)}}
   \parbox[c][1.3in][t]{0.49\textwidth}{\includegraphics[width=.95\hsize, trim=0 0 0 2cm, clip]{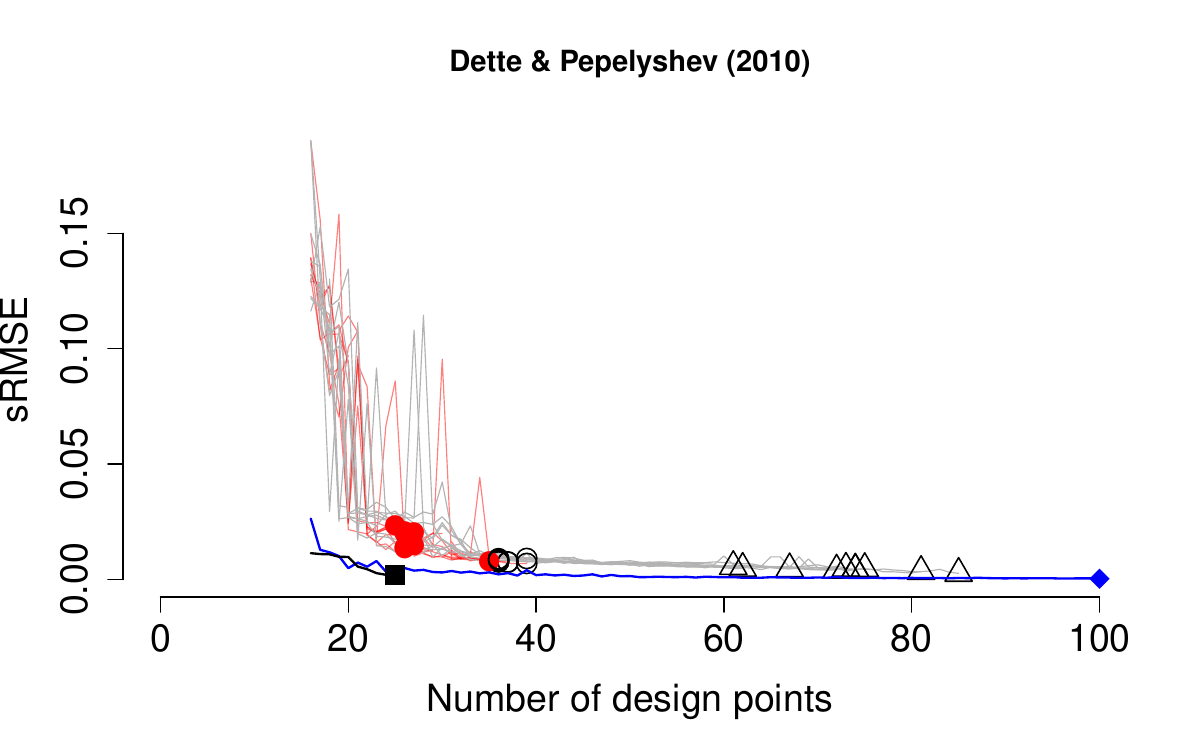}}%
   \parbox[c][1.3in][t]{0.49\textwidth}{\includegraphics[width=.95\hsize, trim=2cm 2cm 2cm 1cm, clip]{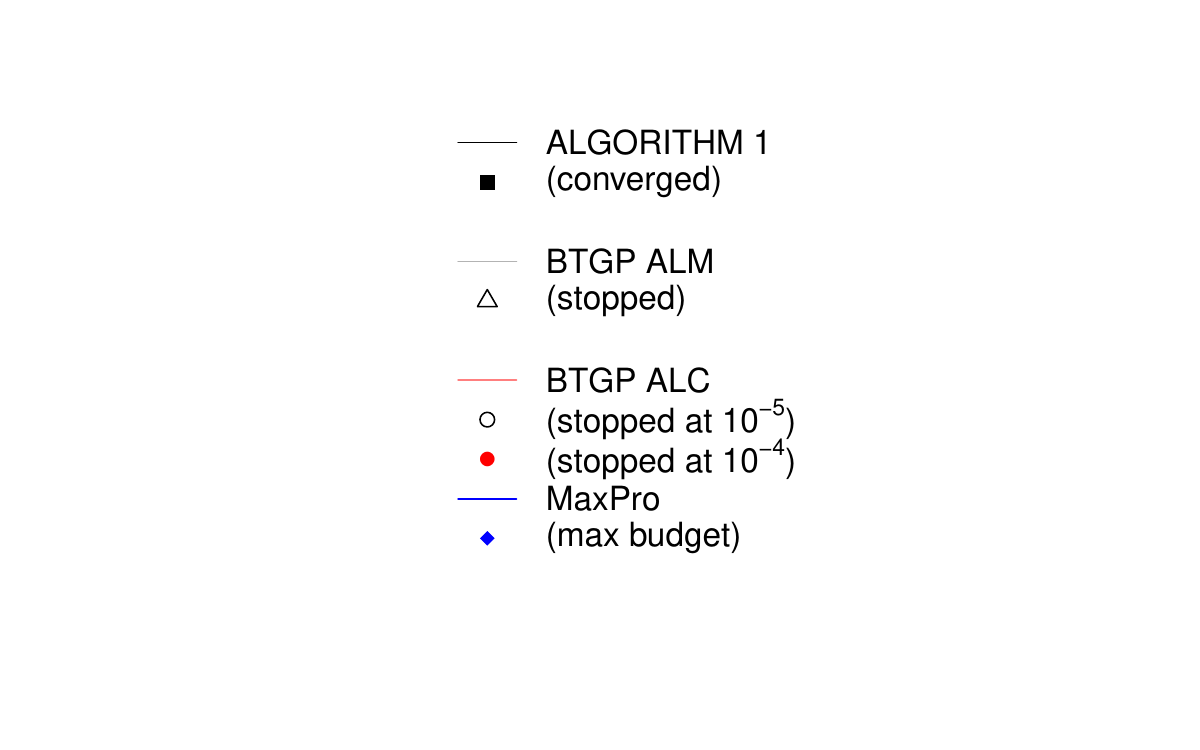}} \\[4ex]
   \parbox[t][.1in][b]{0.49\textwidth}{\subcaption{Dette \& Pepelyshev (2010)}}
\caption{\textbf{Scaled RMSE (sRMSE) with various algorithms for the seven benchmark simulators.}  ALC | ALM: active learning with the Cohn | MacKay criterion, respectively; BTGP: Bayesian treed Gaussian Processes; MaxPro: MaxPro (at the maximum budget)    
\label{fig:benchmarks_sRMSE}}
\end{figure}

Figure~\ref{fig:benchmarks_sRMSE} describes the evolution of the sRMSE with the number of design points for MaxPro, Bayesian treed GPs with the ALC and ALM criteria (10 re-runs for each), and ALGORITHM 1; the corresponding plots for sMAX are qualitatively similar (see Appendix Figure~\ref{fig:benchmarks_sMAX}).

Observe that the lower a line tracks in the plots, the smaller the sRMSE for a given number of design points and the more efficient the corresponding algorithm in approximating the simulator: The scaled RMSE trajectory for ALGORITHM 1 (all benchmarks) and MaxPro (all benchmarks but (f)) are close to or on an efficiency frontier. The trajectories for the Bayesian treed GPs also attain low sRMSE and sMAX, but have variability. An apparent advantage of the active learning designs is that they terminate on their own after $n \leq B$ design points; $n$ ranges from 14 to 82 for ALGORITHM 1 and from 17 to 100 for the various Bayesian treed GPs in Table~\ref{tab:benchmark_results}. By contrast for MaxPro, and for any emulator-free design, the analyst should prespecify $B$. Choosing $B=20$, for example, would result in sRMSE > 5\% with MaxPro for functions (c) and (f). 

\begin{figure} [t]
\captionsetup[subfigure]{margin={0cm,0cm}}
  \parbox[c][2in][t]{0.49\textwidth}{\centering \includegraphics[width=.8\hsize]{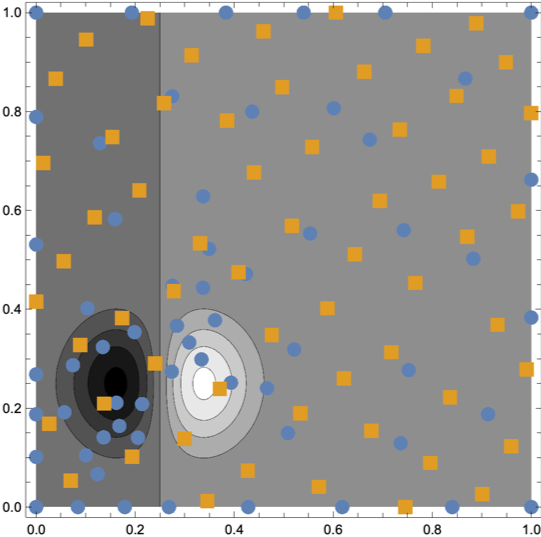}}%
  \parbox[c][2in][t]{0.49\textwidth}{\centering \includegraphics[width=.8\hsize]{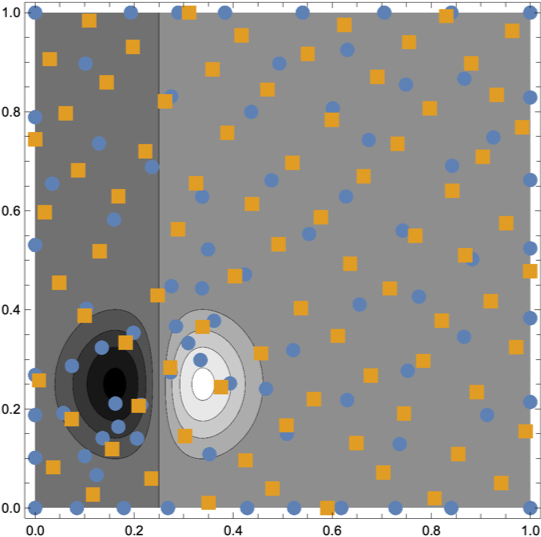}}\\[4ex]
  \parbox[t][.1in][c]{0.49\textwidth}{\subcaption{60 design points}}%
  \parbox[t][.1in][c]{0.49\textwidth}{\subcaption{82 design points}}
\caption{\label{fig:benchmark_ALG1_MaxPro}%
\textbf{Design point distribution with MaxPro and ALGORITHM 1 for benchmark function (f)} Shown is a contour plot for function (f), Gramacy \& Lee (2008). 
Darker shades correspond to the valley and lighter shades correspond to the peak. Yellow squares and blue circles are design points with MaxPro and ALGORITHM 1, respectively. In panel (a), ALGORITHM 1 placed its 60 points with higher density of design points in the lower left quarter of the input space, where the function is non-smooth, and lower density elsewhere. Panel (b) shows the design point distribution when ALGORITHM 1 stopped. It has increased the density of design points primarily in the part of the input space where the function is smooth.  In both panels, MaxPro fills the input space uniformly. }
\end{figure}

The varying smoothness of benchmark (f) over its input domain may explain why the sRMSE trajectory for MaxPro is not on the efficient frontier. The left panel in Figure~\ref{fig:benchmark_ALG1_MaxPro} illustrates that, by 60 design points, ALGORITHM 1 has placed higher density of design points where the function changes most, while MaxPro explores the input space uniformly.

For the Bayesian treed GPs, the sRMSE trajectories and the number of design points at which they converge vary across the 10 re-runs for each benchmark function (Table~\ref{tab:benchmark_results}). In part this may be because they are fit with Markov Chain Monte Carlo and have more parameters than the GPs in ALGORITHM 1: Regularly (here, every 10 design points) the treed GPs examine alternative partitionings of the input space, and fit stationary GPs with noise parameters in each partition. Overall, the treed GPs achieve low sRMSE and sMAX, but on average converge at a higher number of design points compared to ALGORITHM 1. For function (f), treed GPs partitioned the input space in 3 to 5 axis-aligned subdomains in 7 out of 10 and 2 out 10 replications using the ALC and ALM criterion, respectively.  Compared with the treed GPs, ALGORITHM 1 is less variable (it is deterministic) and required on average a smaller number of design points for most functions (all, except for (d) and (e)).

Finally, although the convergence thresholds for the ALM and the ALC convergence criteria were set similarly across the seven benchmarks, they turned out to be either too stringent or too lenient for different functions. For example, some or all of the 10 ALC experiments did not converge before exhausting the budget for functions (a), (b), (c), (e) and (f) using the stricter threshold of a change in ALC $<10^{-5}$. Using the more lenient threshold (change in ALC $<10^{-4}$) results in sRMSE in excess of 5\% for some experiments in functions (c) and (f). With the ALM criterion, 2 experiments fail to converge before exhausting the budget in (f), and all experiments in (g) required up to 3 times the number of design points compared to other algorithms (Table~\ref{tab:benchmark_results}). 

\begin{table}
\rowcolors{1}{}{}
\caption{\textbf{Results with active learning design algorithms for seven benchmark functions}}
\label{tab:benchmark_results}
\resizebox{\textwidth}{!}{%
\begin{tabular}{llllll}
\cline{1-6} 
\textbf{Benchmark function}                  & $k$ & $N$ \textbf{(converged)}            & $n$                    & \textbf{sMAX (\%)}                                          & \textbf{sRMSE (\%)}                                         \\
                           &   &  \multicolumn{4}{l}{\textbf{ALGORITHM 1}} \\
(a) Santner (2003)             & 1 & 1 (1)                    & 14                   & 0.6                                                      & 0.1                                                      \\
(b) Higdon (2002)              & 1 & 1 (1)                    & 19                   & 1.2                                                      & 0.4                                                      \\
(c) Gramacy \& Lee (2012)      & 1 & 1 (1)                    & 38                   & 1.7                                                      & 0.2                                                      \\
(d) Lim (2002)                 & 2 & 1 (1)                    & 31                   & 0.7                                                      & 0.2                                                      \\
(e) Branin-Hoo-Picheny (2012)  & 2 & 1 (1)                    & 37                   & 1.7                                                      & 0.4                                                      \\
(f) Gramacy \& Lee (2008)      & 2 & 1 (1)                    & 82                   & 5.6                                                      & 0.6                                                      \\
(g) Dette \& Pepelyshev (2010) & 3 & 1 (1)                    & 25                   & 0.8                                                      & 0.2                                                      \\
                           &   &  \multicolumn{4}{l}{\textbf{BTGP, active learning with MacKay's criterion (ALM)}~\cite{mackay1992} } \\
(a) Santner (2003)             & 1 & 10 (10)                  & 19.5 {[}17 to 24{]}  & \textless{}0.05 {[}\textless{}0.05 to \textless{}0.05{]} & \textless{}0.05 {[}\textless{}0.05 to \textless{}0.05{]} \\
(b) Higdon (2002)              & 1 & 10 (10)                  & 22.5 {[}21 to 25{]}  & \textless{}0.05 {[}\textless{}0.05 to \textless{}0.05{]} & \textless{}0.05 {[}\textless{}0.05 to \textless{}0.05{]} \\
(c) Gramacy \& Lee (2012)      & 1 & 10 (10)                  & 58.5 {[}45 to 70{]}  & \textless{}0.05 {[}\textless{}0.05 to 0.1{]}             & \textless{}0.05 {[}\textless{}0.05 to \textless{}0.05{]} \\
(d) Lim (2002)                 & 2 & 10 (10)                  & 28.5 {[}27 to 33{]}  & 0.1 {[}\textless{}0.05 to 0.2{]}                         & \textless{}0.05 {[}\textless{}0.05 to \textless{}0.05{]} \\
(e) Branin-Hoo-Picheny (2012)  & 2 & 10 (10)                  & 39 {[}38 to 47{]}    & 0.2 {[}0.1 to 0.4{]}                                     & 0.1 {[}\textless{}0.05 to 0.1{]}                         \\
(f) Gramacy \& Lee (2008)      & 2 & 10 (2)                   & 100 {[}16 to 100{]}  & 2.2 {[}0.6 to 52.6{]}                                    & 0.1 {[}0.1 to 9{]}                                       \\
(g) Dette \& Pepelyshev (2010) & 3 & 10 (10)                  & 73.5 {[}61 to 85{]}  & 1.5 {[}1 to 1.9{]}                                       & 0.5 {[}0.3 to 0.6{]}                                     \\
                           &   & \multicolumn{4}{l}{\textbf{BTGP, active learning with Cohn's criterion (ALC: stricter cutoff, $10^{-5}$)}~\cite{cohn1994}} \\
(a) Santner (2003)             & 1 & 10 (0)                   & 100 {[}100 to 100{]} & 0.1 {[}\textless{}0.05 to 0.2{]}                         & \textless{}0.05 {[}\textless{}0.05 to 0.1{]}             \\
(b) Higdon (2002)              & 1 & 10 (2)                   & 100 {[}22 to 100{]}  & \textless{}0.05 {[}\textless{}0.05 to 0.3{]}             & \textless{}0.05 {[}\textless{}0.05 to 0.1{]}             \\
(c) Gramacy \& Lee (2012)      & 1 & 10 (0)                   & 100 {[}100 to 100{]} & 3.5 {[}0.1 to 14.8{]}                                     & 0.8 {[}\textless{}0.05 to 4.9{]}               \\
(d) Lim (2002)                 & 2 & 10 (10)                  & 22.5 {[}21 to 29{]}  & 0.5 {[}0.1 to 0.9{]}                                     & 0.1 {[}\textless{}0.05 to 0.2{]}                         \\
(e) Branin-Hoo-Picheny (2012)  & 2 & 10 (9)                   & 33 {[}32 to 100{]}   & 0.8 {[}0.1 to 1.1{]}                                     & 0.2 {[}\textless{}0.05 to 0.3{]}                         \\
(f) Gramacy \& Lee (2008)      & 2 & 10 (8)                   & 98.5 {[}87 to 100{]}   & 3.1 {[}1.7 to 9.9{]}                                      & 0.2 {[}0.1 to 0.7{]}                                     \\
(g) Dette \& Pepelyshev (2010) & 3 & 10 (10)                  & 36 {[}36 to 39{]}    & 3.1 {[}2.5 to 4.1{]}                                     & 0.8 {[}0.7 to 0.9{]}                                     \\
                           &   & \multicolumn{4}{l}{\textbf{BTGP, active learning with Cohn's criterion (ALC: more lenient cutoff, $10^{-4}$)}~\cite{cohn1994}} \\
(a) Santner (2003)             & 1 & 10 (1)                   & 100 {[}92 to 100{]}  & 0.1 {[}\textless{}0.05 to 0.3{]}                         & \textless{}0.05 {[}\textless{}0.05 to 0.1{]}             \\
(b) Higdon (2002)              & 1 & 10 (10)                  & 20.5 {[}20 to 23{]}  & \textless{}0.05 {[}\textless{}0.05 to 0.1{]}             & \textless{}0.05 {[}\textless{}0.05 to \textless{}0.05{]} \\
(c) Gramacy \& Lee (2012)      & 1 & 10 (10)                  & 20 {[}18 to 23{]}    & 16 {[}14.4 to 17.6{]}                                    & 5.7 {[}5.6 to 6.2{]}                                     \\
(d) Lim (2002)                 & 2 & 10 (10)                  & 22 {[}20 to 29{]}    & 0.7 {[}0.1 to 1.6{]}                                     & 0.2 {[}\textless{}0.05 to 0.4{]}                         \\
(e) Branin-Hoo-Picheny (2012)  & 2 & 10 (10)                  & 26 {[}22 to 30{]}    & 7.7 {[}3 to 9.4{]}                                       & 2.7 {[}0.5 to 3.1{]}                                     \\
(f) Gramacy \& Lee (2008)      & 2 & 10 (10)                  & 74 {[}58 to 81{]}    & 14.1 {[}4.1 to 48.5{]}                                   & 1.6 {[}0.5 to 8.5{]}                                     \\
(g) Dette \& Pepelyshev (2010) & 3 & 10 (10)                  & 26.5 {[}25 to 35{]}  & 5 {[}3 to 8.1{]}                                         & 1.4 {[}0.8 to 2.3{]}   \\           \cline{1-6}                       
\end{tabular}%
}
\begin{tablenotes}
\item  BTGP: Bayesian treed Gaussian Processes.
\item $k$: input dimensions; $n$: number of design points when the algorithm stops;  $N$ (converged): number of emulators (number converged within 100 iterations); sMAX|sRMSE: scaled MAX|RMSE normalized so that the maximum possible value is 1. In the emulator-free design (MaxPro) the whole design budget is used -- not shown here. 
\end{tablenotes}
\end{table}

\subsection{Results with the PSAPC simulator}
\begin{figure} [t!]
\captionsetup[subfigure]{margin={0cm,0cm}}
  \parbox[c][1.2in][t]{0.49\textwidth}{\includegraphics[width=.95\hsize, trim=0 0 0 1.95cm, clip]{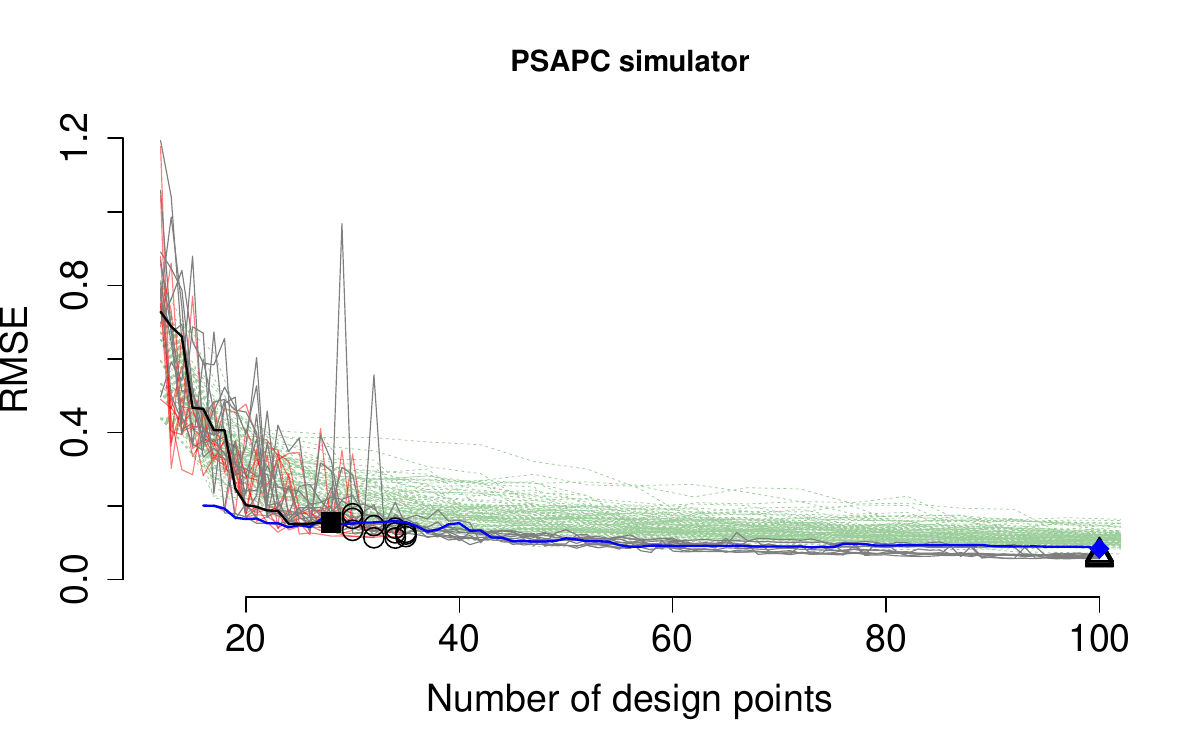}}%
  \parbox[c][1.2in][t]{0.49\textwidth}{\includegraphics[width=.95\hsize, trim=0 0 0 1.95cm, clip]{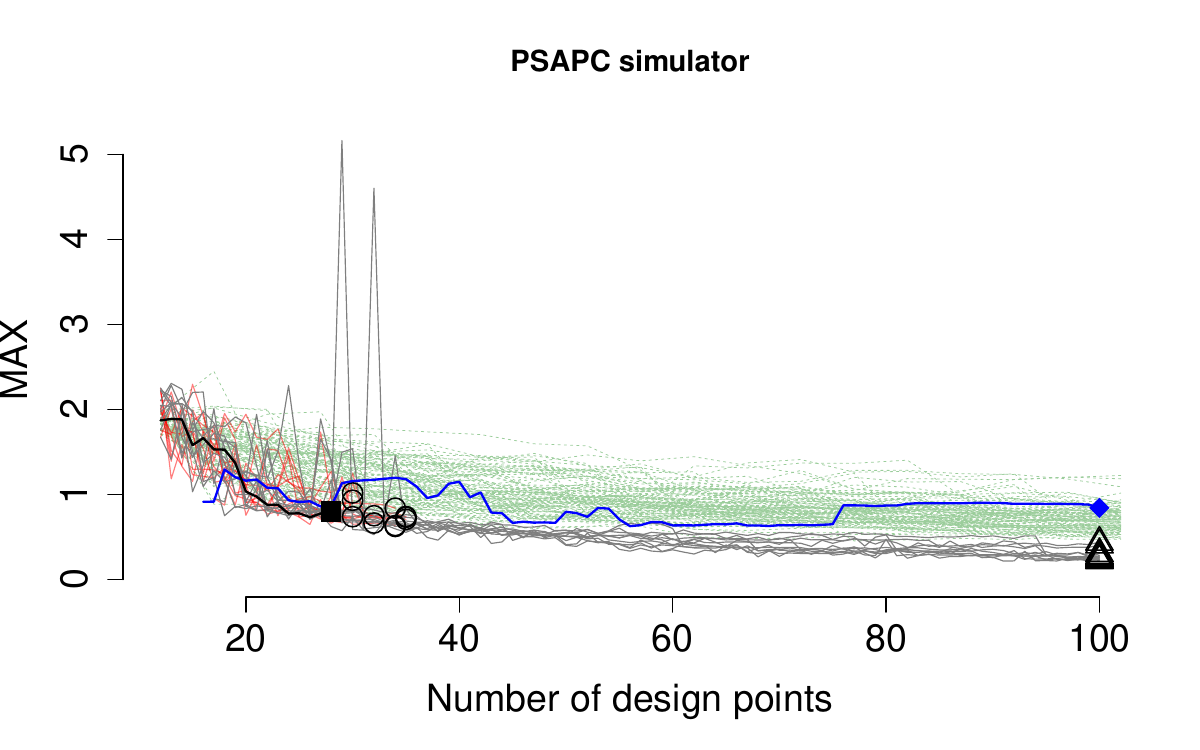}}\\[3ex]
  \parbox[t][.2in][c]{0.49\textwidth}{\subcaption{RMSE, $n_0=12$}}
  \parbox[t][.2in][c]{0.49\textwidth}{\subcaption{MAX, $n_0=12$}}\\[3ex]
  \parbox[c][1.2in][t]{0.49\textwidth}{\includegraphics[width=.95\hsize, trim=0 0 0 1.95cm, clip]{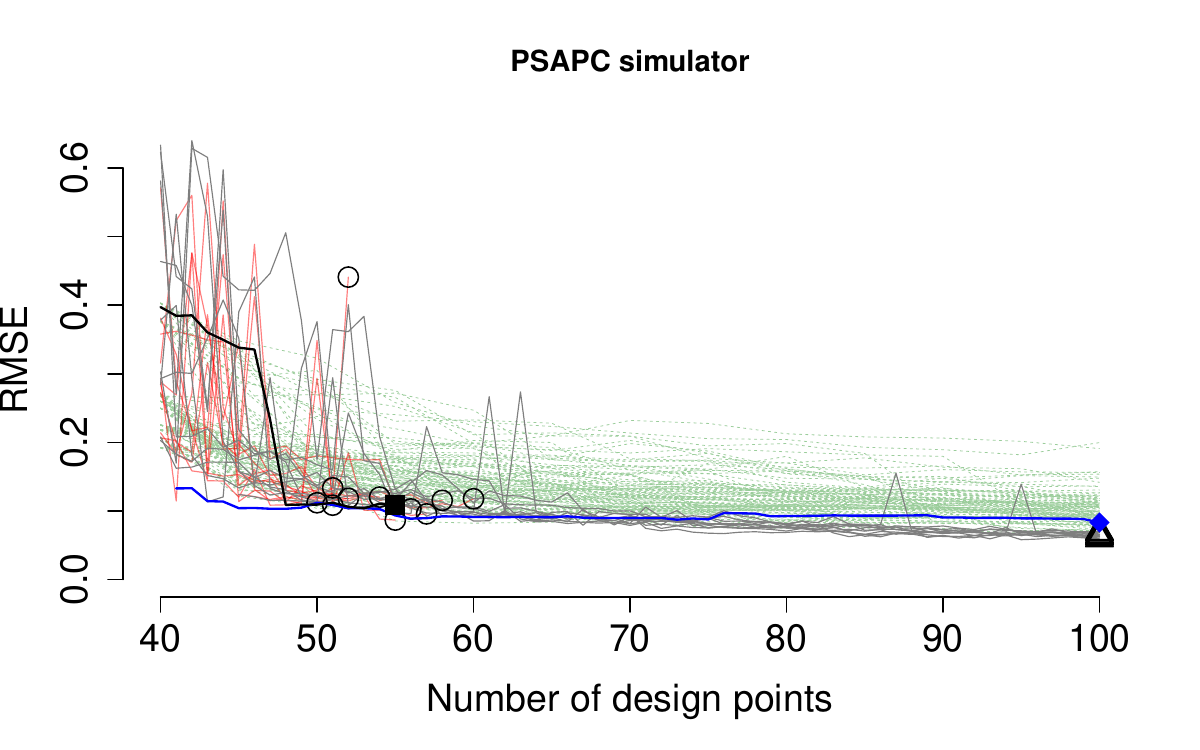}}%
  \parbox[c][1.2in][t]{0.49\textwidth}{\includegraphics[width=.95\hsize, trim=0 0 0 1.95cm, clip]{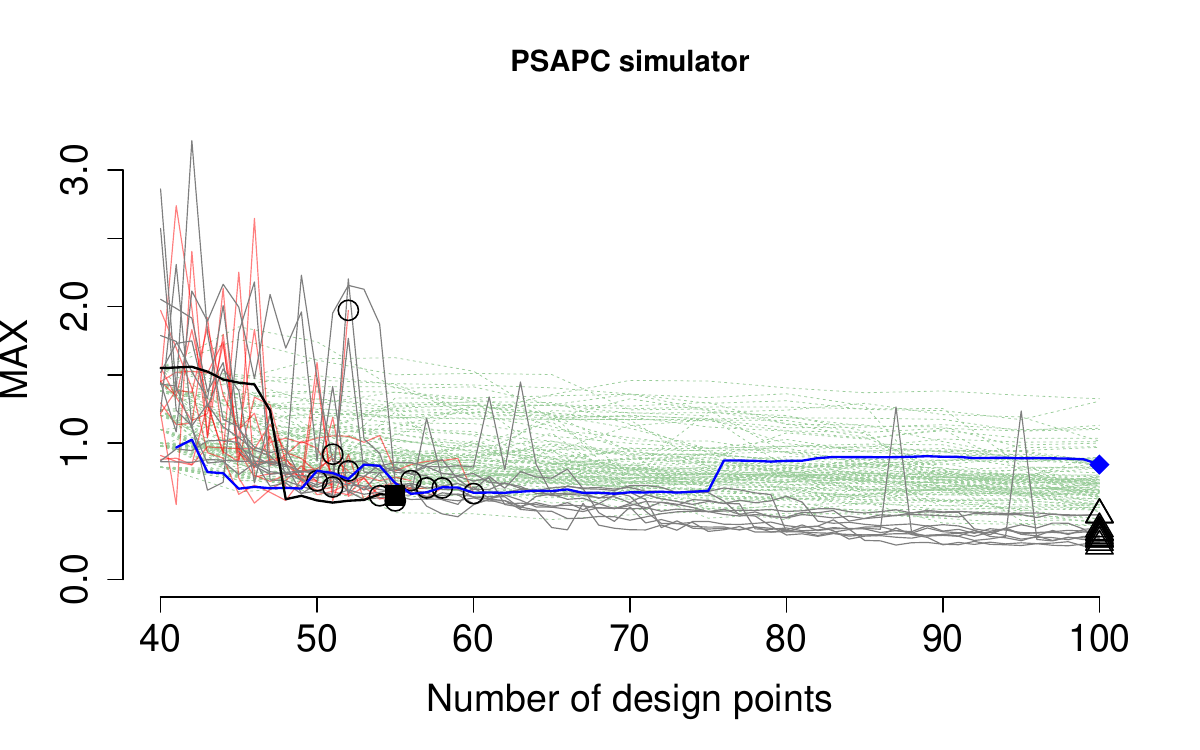}}\\[3ex]
  \parbox[t][.2in][c]{0.49\textwidth}{\subcaption{RMSE, $n_0=40$}}
  \parbox[t][.2in][c]{0.49\textwidth}{\subcaption{MAX, $n_0=40$}}\\[3ex]
  \parbox[c][0.3in][c]{0.95\textwidth}{\includegraphics[width=.95\hsize, trim=0 6cm 0 5cm, clip]{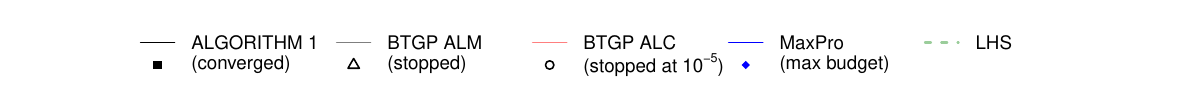}}%
\caption{\textbf{Performance of PSAPC emulators with $k=4$.} RMSE and MAX are measured in quality adjusted days. RMSE and MAX smaller that 1 quality adjusted day represent good performance for the application. The single line for ALGORITHM 1 corresponds to a seeding design whose interior vertices are selected with MaxPro. For visibility we do not show the 10 ALGORITHM 1 trajectories with random seeding designs. The individual lines are hard to see, but they convey the variability of the algorithms. 
\label{fig:psapc_results}}
\end{figure}

Figure~\ref{fig:psapc_results} compares the performance of the five design algorithms for $k=4$, which is the most challenging case. Additional results for $k=1$ through $4$ are shown elsewhere~\cite{ellis2018}.  All algorithms achieved RMSE smaller than 0.5 and MAX smaller than 2 quality adjusted days.  

Of the active learning algorithms, the  treed GPs with the ALM criterion did not converge before exhausting the budget. The convergence threshold was set equal to $T_{SE}=0.5$ quality adjusted days used by ALGORITHM 1, but it was ostensibly too stringent. \footnote{Some difficulty in selecting convergence thresholds for the ALM and ALC criteria was also noted in the benchmark experiments.} Even so, at the maximum budget (and almost throughout the curve), the treed GPs with the ALM criterion achieved lower RMSE and MAX compared to the random LHS classifiers and MaxPro. 

ALGORITHM 1 and treed GPs with the ALC criterion achieved RMSE <0.20 and MAX<1.02 quality adjusted life days with only 28--36 design points, when started with a seeding set of 12 design points. Performance was comparable when the algorithms were started with a seeding set of 40 design points and convergence was achieved between 50 and 60 design points. Performance metrics remained similar, with the exception of one replicate in which a treed GP with the ALC criterion converged having RMSE=0.44 and MAX=1.97 quality adjusted days.  In the ALGORITHM 1 experiments there was little variation in the number of design points needed and the RMSE and MAX achieved over the alternative seeding designs. Treed GPs with ALC showed somewhat larger variability. In all experiments with treed GPs, the models ended up learning two axis-aligned partitions for the whole input space, with the exception of one replicate with the ALM criterion that learned three partitions.

\begin{table}[ht]
\rowcolors{1}{}{}
\centering
\caption{\textbf{Overview of PSAPC emulator performance for $k=4$.} The symbol $n$ is the number of design points used in each design algorithm. Results for the MAX and RMSE are the median [range] for emulators developed with 10 different initial seeds (ALGORITHM 1, BTGP ALM, and BTGP ALC) and 100 emulators trained with Random LHS (10 replications starting from the same 10 random initial sets).
\label{tab:PSAPC_results}}
\resizebox{\textwidth}{!}{%
\begin{tabular}{lcccccc}
\cline{1-7} 
\textbf{Design algorithm} & \multicolumn{3}{c}{$n_0=12$} & \multicolumn{3}{c}{$n_0=40$} \\ \cline{2-7}
 & $n$ & MAX & RMSE & $n$ & MAX & RMSE \\ \hline 
\textit{Emulator-free designs} &  &  &  &  &  &  \\
$\ \ $MaxPro & 100 & 0.84 & 0.08 & 100 & 0.84 & 0.08 \\
$\ \ $Random LHS (100 emulators: 10 replicates $\times$\\
$\ \ \ \ \ \ \ $ 10 random starting sets) & 102 & 0.71 [0.47, 1.22] & 0.11 [0.07, 0.16] & 100 & 0.70 [0.39, 1.32] & 0.10 [0.07, 0.20] \\
\textit{Emulator-based designs} &  &  &  &  &  &  \\
$\ \ $ALGORITHM 1 (MaxPro starting set) & 30 & 0.54 & 0.13 & 51 & 0.69 & 0.09 \\
$\ \ $ALGORITHM 1 (10 random starting sets) & 32 [28, 36] & 0.63 [0.33, 0.91] & 0.14 [0.11, 0.20] & 52 [50, 61] & 0.60 [0.28, 0.76] & 0.11 [0.07, 0.14] \\
$\ \ $treed GPs ALM (MaxPro starting set) & 100 & 0.30 & 0.06 & 100 & 0.47 & 0.06 \\
$\ \ $treed GPs ALM (10 random starting sets) & 100 [100, 100] & 0.26 [0.22, 0.44] & 0.06 [0.06, 0.07] & 100 [100, 100] & 0.32 [0.24, 0.48] & 0.06 [0.06, 0.07] \\
$\ \ $treed GPs ALC (MaxPro starting set) & 30 & 0.74 & 0.13 & 50 & 0.72 & 0.11 \\
$\ \ $treed GPs ALC (10 random starting sets) & 34 [30, 35] & 0.74 [0.62, 1.02] & 0.13 [0.11, 0.18] & 54 [51, 60] & 0.67 [0.58, 1.97] & 0.12 [0.09, 0.44] \\
\hline 
\end{tabular}%
}
\begin{tablenotes}
\begin{footnotesize}
\item ALC|ALM: Active learning with the Cohn~\cite{cohn1994}|MacKay~\cite{mackay1992} criterion; LHS: Latin hypercube sampling
\end{footnotesize}
\end{tablenotes}

\end{table}

\vspace{0.2in}
\section{Discussion}\label{sec:remarks}

In simulators with varying smoothness over the input domain, active learning algorithms resulted in emulators with smaller RMSE and MAX for the same number of design points. In all other cases, all algorithms performed comparably. The proposed algorithm attained satisfactory performance in all analyses, had smaller variability than the treed GPs (it is deterministic), and, on average, had similar or better performance as the treed GPs in 6 out of 7 benchmark functions and in the prostate cancer model.

\subsection{Interpretation of findings}\label{sec:interpretation}

In the empirical analyses with benchmark functions (Figure~\ref{fig:benchmarks_sRMSE}, and Appendix Figure~\ref{fig:benchmarks_sMAX}) all algorithms' trajectories reduced the sRMSE and sMAX with increasing number of design points. 

For benchmark functions whose smoothness does not change dramatically over the input domain, (all except (c) and (f) in Figure~\ref{fig:benchmark_functions}) algorithms that optimize geometric criteria (here MaxPro) are on or near the efficient frontier. \emph{An advantage of the active learning algorithms may be that they can terminate at some $n<B$}, if $B$ is large enough. For designs that implement geometric criteria, $B$ must be pre-specified to match the simulator at hand, but this is not easy to do even when one has detailed information about the simulator.  In the examples, $B<20$ would attain sRMSE and sMAX less than 1\%, for functions (a) and (b), but over 5\% for functions (c) and (f). ``Rules of thumb'' based on simulation testing suggest that trying about 100 points may be sufficient for a few input variables~\cite{Helton2006, Meckesheimer2001}, which would be too conservative for all our benchmark functions. We believe (but have not proven) that for smooth simulators, ALGORITHM 1 will always terminate, because GPs converge to the underlying simulator with increasing number of design points~\cite{Rasmussen2006}.

For benchmark functions such as (c) and (f), whose smoothness varies greatly across the input domain, \emph{active learning algorithms may achieve higher density of design points in the less smooth regions} (Figures~\ref{Fig_AL}(i) and~\ref{fig:benchmark_ALG1_MaxPro}), which is preferable to uniformly filling the input space. (If the emulator is biased, however, this will not be achieved).

Among the active learning algorithms we compared, ALGORITHM 1 is remarkable because it is \emph{deterministic in its evolution} and its sRMSE and sMAX trajectories are close or on the efficient frontier and have smaller variability than those of the treed GPs for the examined functions. Large variability in the error descent trajectories is probably undesirable for expensive simulators.

In the analyses, although the convergence criteria for the Cohn or MacKay criteria were set similarly in the all the benchmark functions and matched those of ALGORITHM 1, we observed variability in achieving convergence. The same criteria were either too lenient or too stringent for different benchmark functions, resulting in all or some repetitions to exhaust the budget or stop too soon.  In practice, we have found that setting convergence criteria is an underappreciated and vexing problem.

Our results are consistent with the literature, which also suggests that sequential designs can approximate simulators with predictive accuracy that is at least comparable to or better than that of conventional space-filling designs \cite{Kleijnen2004, Jin2001}. The efficiency of active learning algorithms is an important consideration for simulators that require substantially more computational resources compared to the PSAPC, such as the physiology-based chronic disease model~\emph{Archimedes}~\cite{schlessinger2002,eddy2003} or detailed modeling of the biomechanics of red blood cells~\cite{tang2017}. In our case, we chose a computationally tractable simulator to quantify the evolution of RMSE and MAX metrics for emulators developed with different designs and to demonstrate computational efficiencies. Such calculations would be impractical with mathematical models that take hours or days, rather than minutes, to evaluate.

\subsection{Applications to cost-effectiveness analysis}\label{sec:CEA}

We believe that this work responds to the call by the Second Panel for Cost-Effectiveness in Health and Medicine for future research on ``practical approaches to efficiently develop well-performing emulators'' of mathematical models in healthcare~\cite{neumann2016,Neumann2018,Sanders2016}. 

Cost-effectiveness-analysis involves two decision-relevant quantities, the marginal expected effectiveness $\Delta E$ and the marginal expected cost $\Delta C$ between two interventions, which suggests the need for multivariate emulation. At least two avenues are possible. First, for a given willingness-to-pay $\lambda$ (the shadow price of an additional effectiveness unit), identifying optimal strategies with a cost effectiveness analysis is strategically equivalent with maximizing the expected net health benefit,  $NHB = \Delta E - \Delta C / \lambda$, which translates costs to effectiveness units, or with maximizing the net monetary benefit $NMB = \lambda \Delta E - \Delta C$, which translates effectiveness to monetary units. This would combine the two decision relevant quantities in a single measure, requiring univariate emulators. The $NHB$ or $NMB$ would be parameterized by $\lambda$, which can be another input to the emulator.

Multivariate output emulators extend the GP model in the appendix to $q>1$ outputs~\cite{Conti2010}, and were proposed early on as a promising avenue~\cite{ohagan1998}. Somewhat surprisingly, empirical~\cite{kleijnen2014} and theoretical~\cite{Gu2016,kyzyurova2017} results suggest that, for the purpose of model output emulation, performance with multivariate emulators does not exceed that of separate univariate emulators. Thus, a practical alternative is to learn separate univariate emulators, one for each quantity of interest.

\subsection{Simulators with noisy observations}

Some clarifying comments are warranted with respect to simulators that yield noisy observations. 
In some cases, some simulator inputs are random variables (e.g., because they have been estimated in finite samples), \emph{the mean outputs will also be random}. These stochastic simulators should be emulated with GPs that assume the observations are noisy (unless the noise is too small for the application). An example is CISNET's breast cancer model \emph{M}, a fully Bayesian model~\cite{berry2006}.

In many other cases, including in our application, \emph{the residual noise is the forward-Monte Carlo error of a numerical integration}. Monte Carlo error can be minimized with various simulation techniques (e.g., using common random numbers in simulations comparing different strategies~\cite{nelson1995}) or simply by running bigger simulations. Many microsimulation-based simulators in health fit this paradigm, including the majority of CISNET models, where random inputs are fixed to a reference baseline value.  It may by interesting to explore emulation strategies for different allocations of a fixed computational budget, e.g., obtain more-noisy observations for a larger number of design points.

Using GPs that allow for observation noise has been advocated on philosophical (e.g., the simulator models a target phenomenon with possibly nontrivial uncertainty~\cite{gramacy2010,OHagan2006}) and practical grounds~\cite{ranjan2011} (e.g., for numerical stability), even for deterministic simulators. We set very modest goals for emulation in that we approximate the simulator rather than the target phenomenon, so we treat observations from deterministic simulators as noise-free.

\subsection{Limitations}\label{sec:limitations} 

Several of the limitations pertain to the fact that we use GPs for emulation, which do not scale well to many inputs, do not extrapolate well, and can themselves be expensive to fit~\cite{Rasmussen2006}.
O'Hagan suggests that that GPs may be practical for up to $50$ input dimensions~\cite{OHagan2006}. If a large number of inputs must be emulated, one should explore whether developing several lower-dimensional emulators suffices. Knowing the  analytic specification of the underlying model could offer insights about the feasibility of this simplification. GPs are interpolators and should not be used to extrapolate outside the input polytope. If such extrapolations are required, other types of emulators should be used, as the emulation problem has different goals that those described in Section~\ref{sec:emulation_goals}.

The computational cost of learning GPs is about $O(n^3)$~\cite{Rasmussen2006} in the number $n$ of design points, and for large $n$, fitting GPs becomes computationally expensive. ALGORITHM 1 requires successively fitting a large number of GPs. However, it fits GPs that have one less or one more design point than an already fitted GP. Re-fitting in a minimally reduced or augmented dataset can be sped up by using previously-estimated parameter values as starting values for the fitting algorithm (as was done here), and with other approximations~\cite{shen2006}.  

GPs trained with active learning are not guaranteed to converge to the simulator~\cite{pronzato2012}, whereas GPs trained with LHS will eventually do so~\cite{mckay1979}. Asymptotic convergence would probably be achieved by modifying step 5 in ALGORITHM 1 to also allow choosing, with some probability, a set of random design points in an ``exploration'' step. However, this remedy might undercut the algorithm's efficiency gains.  

\subsection{Extensions }\label{sec:extensions} 

ALGORITHM 1 is a meta-heuristic and can accommodate many different emulator models. For example, it can be modified by substituting GPs that model noisy observations to emulate stochastic simulators.
Some care would be needed to avoid setting too stringent a resampling or standard error threshold in the stopping criteria. For example, if the $T_{SE}$ threshold is smaller than the simulator's prediction uncertainty, the algorithm would not terminate.  
ALGORITHM 1 can also use other types of emulators, including regression or neural network learners, although we have not examined this case in an example. 
An additional application is the case of \emph{linked emulators}, that are used to approximate systems of computer simulators, or equivalently, distinct modules of a modular simulator, where the results of one module are used by other modules to yield the output of the overall simulator~\cite{kyzyurova2017,kyzyurova2018}. Focused theoretical analyses and limited empirical applications suggest that one should be able to fit separate emulators for each simulator module with ALGORITHM 1.

For emulating multivariate simulator outputs a multivariate GP can be used, which can also model correlations between the multiple outputs \cite{Rasmussen2006,Conti2010,Genton2015}. If multivariate emulators are to be used, the stopping criteria for ALGORITHM 1 would also need to be modified, e.g., $T_{resample}$ and $T_{SE}$ should be be met for each output, or for some function of the outputs. However, in practice, for a large class of models, there appears to be little gain attempting to learn multivariate emulators compared to several univariate ones~\cite{kleijnen2014,Gu2016,kyzyurova2017}.

As described, ALGORITHM 1 is better suited to approximating a function over all its input domain, rather than finding optima. Even so, for the PSAPC model, the emulators learned with ALGORITHM 1 had MAX<1 quality adjusted day, which approximates the PSAPC model sufficiently well. For example, emulators with a MAX=0.5 (ALGORITHM 1 starting from a MaxPro-based design of 12 points had MAX=0.54) would suffice to correctly rank the 17 expert-generated strategies for annual or biennial screening in Gulati et al. 2013 that can be evaluated with our version of PSAPC~\cite{Gulati2013}, many of which differ by almost a quality adjusted month.
However, to efficiently search the space of practically-implementable policies, it may be better to use emulator-based algorithms that aim to approximate the simulator \emph{in the neighborhood of an optimum, rather throughout the input domain}. Such an algorithm is described in Ellis 2018~\cite{ellis2018}.

Finally, in applications where it is critical to train emulators to a target accuracy or better, some more effort is required. By its construction, ALGORITHM 1 provides some information on the accuracy of the emulator at each iteration, because it uses a resampling scheme to identify the candidate points. Thus, at each iteration, one can examine an RMSE-like quantity such as $\sqrt {\mathbb{E}_i \ [f^*_{(-i)}(\bm{x}_i) - f(\bm{x}_i)]^2}$ to gauge the accuracy of the resulting emulators, and perhaps, make the convergence thresholds more stringent~\cite{ellis2018}. However, more generally, one can gauge the accuracy of the developed emulator by generating additional design points and estimating the RMSE of the emulator to decide whether to further tighten the algorithm's stopping criteria. Adding the additional points to the design, on average, will result in emulators with even better RMSE. Ellis 2018 proposes another meta-heuristic algorithm that can train emulators to a target accuracy, albeit at an increased computational cost~\cite{ellis2018}.

\section*{acknowledgements}
We thank Dr. Roman Gulati for providing the PSAPC model source code and for answering technical questions about the implementation.  

\section*{conflict of interest}
We have no conflicts to declare. 


\bibliography{references}

\newpage
\appendix
\section*{Appendices}
\renewcommand{\thesubsection}{\Alph{subsection}}

\renewcommand{\thetable}{\thesubsection.\arabic{table}}
\setcounter{table}{0}

\renewcommand{\thefigure}{\thesubsection.\arabic{figure}}
\setcounter{figure}{0} 

\subsection{Review of Gaussian Processes}\label{app:GPs}
A Gaussian Process is defined as \emph{a collection of random variables, any finite number of which have a joint normal distribution}~\cite{Rasmussen2006}.
A Gaussian Process $f^*(\bm{x})$ is specified by its mean function 
$$m(\bm{x}) = \mathbb{E}[f^*(\bm{x})]$$
and covariance function 
$$\mathrm{cov} \big(f^*(\bm{x}_i), f^*(\bm{x}_j) \big) = k(f(\bm{x}_i), f(\bm{x}_j)) = \mathbb{E}\Big [ \big( f^*(\bm{x}_i)-m(\bm{x}_i) \big ) \big (f^*(\bm{x}_j)-m(\bm{x}_j) \big ) \Big ]. $$ 
We write the notation $f^*(\cdot) \sim \mathcal{GP} \big ( m(\cdot), k(\cdot,\cdot) \big)$.  

In practice, a simulator $f(\cdot)$ is evaluated $n$ times to obtain a set of design points, $\mathcal{D}_{n} = \{(\bm{x}, f(\bm{x})): \bm{x} \in \mathcal{X}_n\}$. Typically, the inputs $\bm{x}$ are transformed so that their domain is the unit hypercube. A Gaussian Process emulator model approximates the simulator $f(\cdot)$ with 
$$f^*(\cdot) = m(\cdot) + \delta(\cdot),$$ 
where $\delta(\cdot) \sim \mathcal{GP} \big( \textbf{0}, k(\cdot, \cdot) \big)$ is a zero-mean Gaussian Process and $\textbf{0}=(0,\dots,0)^T$. 

Commonly, the mean function is a constant model $m(\bm{x})=\mu$, but it may also be a function of the input vectors, e.g., a linear component $m(\cdot) = \bm{h}(\cdot)^T\bm{\beta}$, where $\bm{\beta}$ is a vector of $p$ coefficients and $\bm{h}(\cdot)$ is a column vector of $p$ known functions of the input vectors. Under this specification, the residuals from the linear regression are modeled by a zero-mean Gaussian Process. Although specifying a non-constant $m(\cdot)$ is not necessary, doing so may increase the smoothness of the fitted Gaussian Process and hence may require fewer design points for a ``good'' model fit \cite{OHagan2006}.

The ``behavior'' of the zero-mean Gaussian Process $\delta(\cdot)$ is determined by the covariance function $k(\cdot, \cdot)$. Many choices for the covariance function exist; see \cite{Rasmussen2006} for a discussion. A common class of functions assumes a stationary covariance process, i.e., that the relationship between any two points depends only on their distance and not their location. Within this class, a typical choice for the covariance function is the squared exponential or radial basis function 
$$k(f(\bm{x}_i), f(\bm{x}_j)) = \sigma^2 \ \exp \big(-(\bm{x}_i - \bm{x}_j)^T \bm{\Theta} (\bm{x}_i - \bm{x}_j) \big) = \sigma^2 \prod_{d=1}^k \exp \big( -\theta_d (x_{di} - x_{dj})^2 \big) = \sigma^2 \ c(f(\bm{x}_i), f(\bm{x}_j)),$$ 
where $\sigma^2$ is the variance of the process, $\bm{\Theta} = \textrm{diag}(\theta_1, \dots, \theta_k)$ a diagonal matrix of $k$ non-negative roughness parameters, and $c(\cdot, \cdot)$ the implied correlation function. Note that $c(f(\bm{x}_i), f(\bm{x}_i))=1$, and that the correlation between $\bm{x}_i, \bm{x}_j$ is positive and decreases with the distance between $\bm{x}_i$ and $\bm{x}_j$~\cite{Kleijnen2000,Rasmussen2006}. The roughness parameters $\bm{\theta}=(\theta_1, \ldots, \theta_k)^T$ determine how quickly the correlation decreases in each dimension. A large value suggests a low correlation for a given dimension, even when two points are close.


Let $\bm{y} = (y_1, \dots, y_n)^T = (f^*(\bm{x}_1), \dots, f^*(\bm{x}_n))^T$ be a vector of observations modeled with a Gaussian Process. From the definition of the Gaussian Process, these observations are modeled as following an $n$-dimensional multivariate normal distribution. The parameters $\bm{\beta}$, $\sigma^2$, and $\bm{\theta}$ of the mean function and the covariance function can be obtained by optimizing the likelihood of the observations~\cite{Kleijnen2008, Rasmussen2006, Macdonald2015}. 
For example, for a constant mean function $m(\bm{x})=\mu$, the mean $\mu$ and the variance $\sigma^2$ of the Gaussian Process can be expressed in closed form as a function of the roughness parameters 
$$\mu = \frac{\textbf{1}^T\bm{R}^{-1}\bm{y}}{\textbf{1}^T\bm{R}^{-1}\textbf{1}}, \textrm{ and}$$
$$\sigma^2 = \frac{(\bm{y} - \textbf{1}\mu)^T\bm{R}^{-1}(\bm{y} - \textbf{1}\mu)}{n},$$
where $\textbf{1}=(1, \dots,1)^T$ and $\bm{R}$ is a $n\times n$ symmetric positive definite correlation matrix in which the correlation between the $i$-th and $j$-th observations is $c(f(\bm{x}_i), f(\bm{x}_j))$. 
The negative profile log likelihood $L_{\bm{\theta}}$ 
$$-2\mathrm{log}(L_{\bm{\theta}})\propto \mathrm{log}(|\bm{R}|) + n\mathrm{log}[(\bm{y}-\textbf{1}\mu)^T\bm{R}^{-1}(\bm{y}-\textbf{1}\mu)]$$ 
is a function of the roughness parameters $\bm{\theta}$ because $\mu, \sigma^2$, and $\bm{R}$ are functions of $\bm{\theta}$. 
In the equation above, $|\bm{R}|$ is the determinant of $\bm{R}$.  Optimizing the likelihood yields estimates $\hat{\bm{\theta}}$ for $\bm{\theta}$, and thus estimates for $\mu, \sigma^2$ and the correlation matrix $\bm{R}$.  

The best linear unbiased predictor of the output value at a new input vector $\tilde{\bm{x}}$ is \cite{Kleijnen2008, Rasmussen2006}
$$f^*(\tilde{\bm{x}}) = {\mu} + \bm{r}^T\bm{R}^{-1}(\bm{y} - \textbf{1}{\mu}),$$ 
where $\bm{r}$ is a $n$-vector of correlations between $\tilde{\bm{x}}$ and the $n$ input vectors from the design points, such that the $i$-th element of $\bm{r}$ is $c(f(\tilde{\bm{x}}), f(\bm{x}_i))$. The prediction error is \cite{Kleijnen2008, Rasmussen2006}
$$s^2(\tilde{\bm{x}}) = \sigma^2 \left[1-\bm{r}^T\bm{R}^{-1}\bm{r}+\frac{(1-\textbf{1}^T\bm{R}^{-1}\bm{r})^2}{\textbf{1}^T\bm{R}^{-1}\textbf{1}} \right].$$ The maximum likelihood estimate $\hat{\bm{\theta}}$ is used to obtain $\bm{r}$ and $\bm{R}$ in the two prediction formulas above.  

Finally, note that estimating GP model parameters involves inverting the correlation matrix $\bm{R}$. If any pair of design points in the input space are close together, $\bm{R}$ may become near-singular, and the fitting procedure may become computationally unstable. To this end, one can substitute $\bm{R}_\delta = \bm{R} + \delta \bm{I}$ for $\bm{R}$ in the formulas above, where $\delta>0$ is a known as the ``nugget'' term, and $\bm{I}$ is the identity matrix. We select the nugget to be as small as practical following the the iterative Tikhonov regularization scheme in~\cite{ranjan2011}. While, generally, small nuggets should have a negligible effect in the fitted GP, addition of the nugget can appreciably affect the overall shape and modes of the GP likelihood~\cite{pepelyshev2010}.

\subsection{Additional information about the simulators}
\subsubsection{Benchmark functions}\label{app:benchmark_specs}
Table ~\ref{tab:benchmarks_specs} shows the specifications of the benchmark functions, their extreme values and their range. All benchmark functions have been linearly transformed from their original version in the respective citations so that their inputs are in $[0, 1]^k$ (except for function (g), whose inputs are in $(0, 1]^3$).
\begin{table}[bt!]
\rowcolors{1}{}{}
\centering
\caption{\textbf{Specification extreme values and range for the seven benchmark functions}.}
\label{tab:benchmarks_specs}
\resizebox{\textwidth}{!}{%
\begin{tabular}{llllll}
\textbf{Benchmark function} & $k$ & \textbf{Maximum} & \textbf{Minimum} & \textbf{Range} & \textbf{Specification} \\
(a) Santner (2003)~\cite{santner2003} & 1 & 1.000 & -0.676 & 1.676 & $\exp{(-1.4 x)} \cos{(3.5 \pi x)}$ \\
(b) Higdon (2002)~\cite{higdon2002} & 1 & 1.142 & -1.142 & 2.284 & $\sin{(2\pi x)}+ 0.2 \sin{(8 \pi x)}$ \\
(c) Gramacy \& Lee (2012)~\cite{gramacy2012} & 1 & 5.063 & -0.896 & 5.959 & $-\sin{(20 \pi x)} / (4 x +1) + (2x-0.5)^4$ \\
(d) Lim (2002)~\cite{lim2002} & 2 & 9.556 & 1 & 8.556 & $9 + 2.5 x_1 - 17.5 x_2 + 2.5 x_1 x_2 + 19 x_2^2 - 7.5 x_1^3 - 2.5 x_1 x_2^2 - 5.5 x_2^4 + x_1^3 x_2^2$ \\
(e) Branin-Hoo-Picheny (2012)~\cite{picheny2013} & 2 & 4.876 & -1.047 & 5.924 & $\frac{1}{51.95} \Big(  \big( 15 x_2 - \frac{5.1}{4 \pi^2}(15 x_1 -5)^2 + \frac{5}{\pi}(15 x_1 -5) -6  \big)^2 + (10 -\frac{5}{4 \pi}) \cos{(15 x_1 -5)} -44.81  \Big)$ \\
(f) Gramacy \& Lee (2008)~\cite{gramacy2008b,gramacy2008} & 2 & 0.429 & -0.429 & 0.858 & $(8 x_1 - 2) \exp{\big(-(8 x_1 - 2)^2 - (8 x_2 - 2)^2) \big)}$ \\
(g) Dette \& Pepelyshev (2010)~\cite{dette2010} & 3 & 1.000 & 0.000 & 1.000 & $2.463019 \Big(\exp{(-\frac{2}{x_1^{1.75}})} +\exp{(-\frac{2}{x_2^{1.5}})} + \exp {(-\frac{2}{x_3^{1.25}})} \Big)$
\end{tabular}%
}
\end{table}


\subsubsection{The PSAPC model}\label{app:psapc_model}
The PSAPC microsimulation model accounts for the relationship between PSA levels, prostate cancer disease progression, and clinical detection~\cite{CISNETpc,Gulati2010}. The model, its estimation approach, its calibration, and its comparison with other prostate cancer models have been described in detail elsewhere~\cite{Gulati2010, Gulati2013, Gulati2014, psapc2009}. Here, we treat the PSAPC model as a ``black box''. 

\begin{figure}[ht]
 \centering
\includegraphics[width=\textwidth]{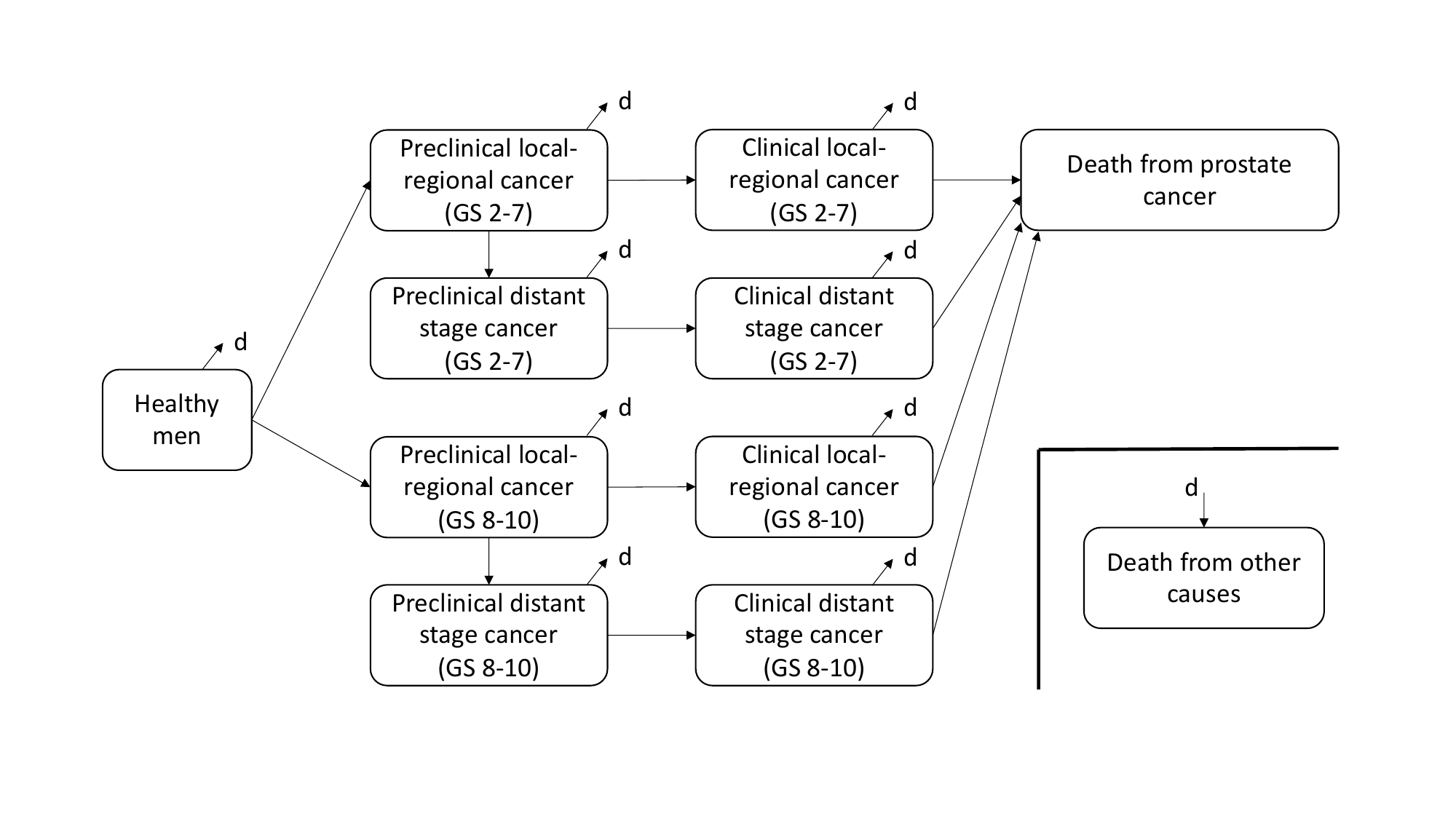}
\caption{\label{fig:GulatiFig}%
PSAPC model natural history of disease progression. Healthy, preclinical, clinical, prostate cancer mortality, and other-cause mortality states in the absence of screening. Rounded rectangle represent the various health states. Arrows between rectangles represent allowable transitions between health states. People develop preclinical local-regional or distal disease, which may manifest clinically. Patients can die of prostate cancer only after they have developed clinical local regional or distal disease. People can die of other causes from any ``alive'' health state. For simplicity, transitions from the ``alive'' health states to ``death from other causes'' are not drawn explicitly, but are depicted by the broken arrows and the letter ``d''. GS: Gleason Score.}
\end{figure}

Figure~\ref{fig:GulatiFig} outlines the PSAPC model. Briefly, simulated healthy men may develop preclinical, local-regional cancer (disease onset). The PSAPC version we are using incorporates disease grade (low=Gleason scores 2-7; high=Gleason scores 8-10) which is determined and fixed upon disease onset. Patients with low- or high-grade, local-regional cancer may progress to distant sites (metastatic spread). Patients with either local-regional or metastatic disease may present with symptoms (clinical detection). Those with a clinically-detectable form of disease may die from prostate cancer (prostate cancer mortality). At any time and any health state in the model, patients may die from other causes (other-cause mortality). Disease progression is related to age or PSA levels. PSA levels are modeled as a function of age and age of disease onset, such that there is a linear changepoint in (log) PSA after disease onset. PSA levels after disease onset differ for those with low versus high-grade disease. Parameters for the age of disease onset, metastatic spread, and clinical detection are estimated from calibration.

In the presence of screening, the simulated individuals with cancer may be identified and treated earlier (i.e., during preclinical state) than without screening. The model assigns each simulated individual a schedule of PSA-based screening tests and biopsies, as determined by the simulated screening strategy. Every time screening occurs, men with PSA levels above the PSA positivity threshold are referred to biopsy, and those with a positive biopsy result are managed with radical prostatectomy, radiation therapy, or active surveillance (i.e., no treatment but continued monitoring)~\cite{CISNETpc,Gulati2010}. 

The PSAPC model projects several outcomes, including the number of screenings, false-positive results, prostate cancer diagnoses, prostate cancer deaths, and life-years gained with PSA-screening versus no screening (i.e., clinical detection only). The model results are presented as the mean number of events or the lifetime probability of each outcome based on the simulated cohorts of men (e.g., 100 million men)~\cite{Gulati2013}.

\subsection{Additional results with benchmark functions}\label{app:benchmark_sMAX}

\begin{figure} []
\captionsetup[subfigure]{margin={0cm,0cm}}
   \parbox[c][1.3in][t]{0.49\textwidth}{\includegraphics[width=.95\hsize, trim=0 0 0 2cm, clip]{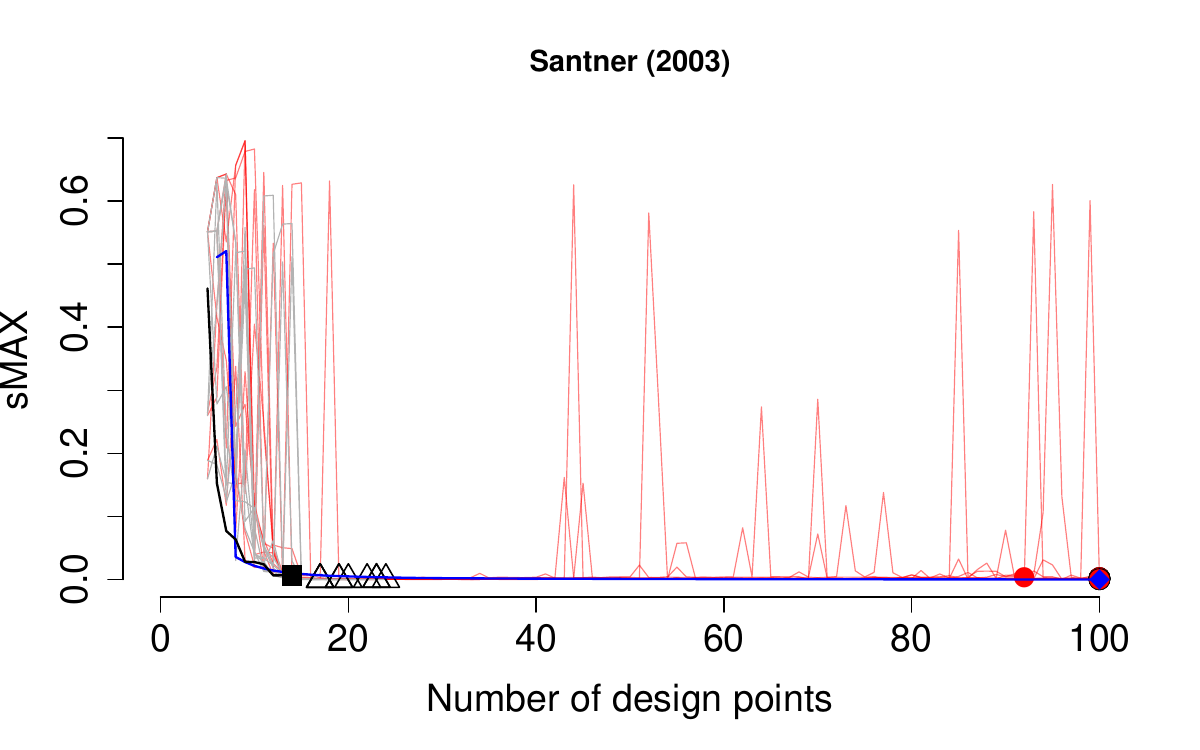}}%
   \parbox[c][1.3in][t]{0.49\textwidth}{\includegraphics[width=.95\hsize, trim=0 0 0 1.73cm, clip]{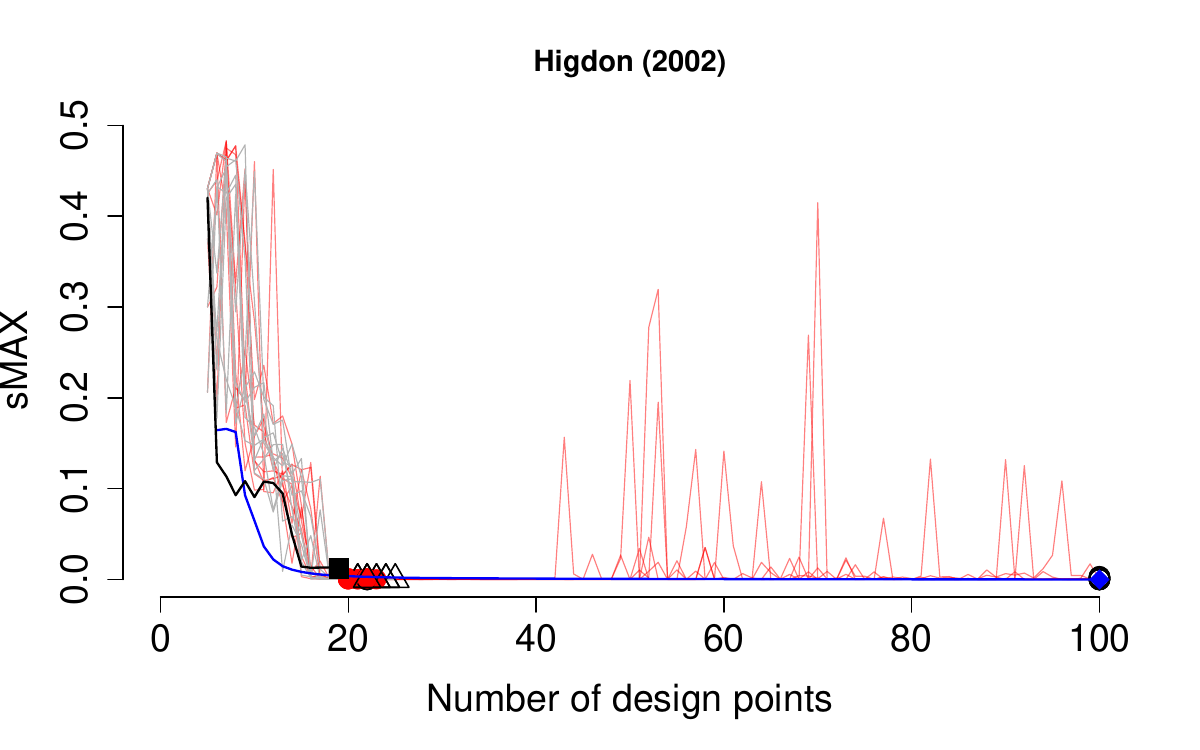}}\\[4ex]
   \parbox[t][.1in][c]{0.49\textwidth}{\subcaption{Santner (2003)}}
   \parbox[t][.1in][c]{0.49\textwidth}{\subcaption{Higdon (2002)}}\\[4ex]
   \parbox[c][1.3in][t]{0.49\textwidth}{\includegraphics[width=.95\hsize, trim=0 0 0 1.8cm, clip]{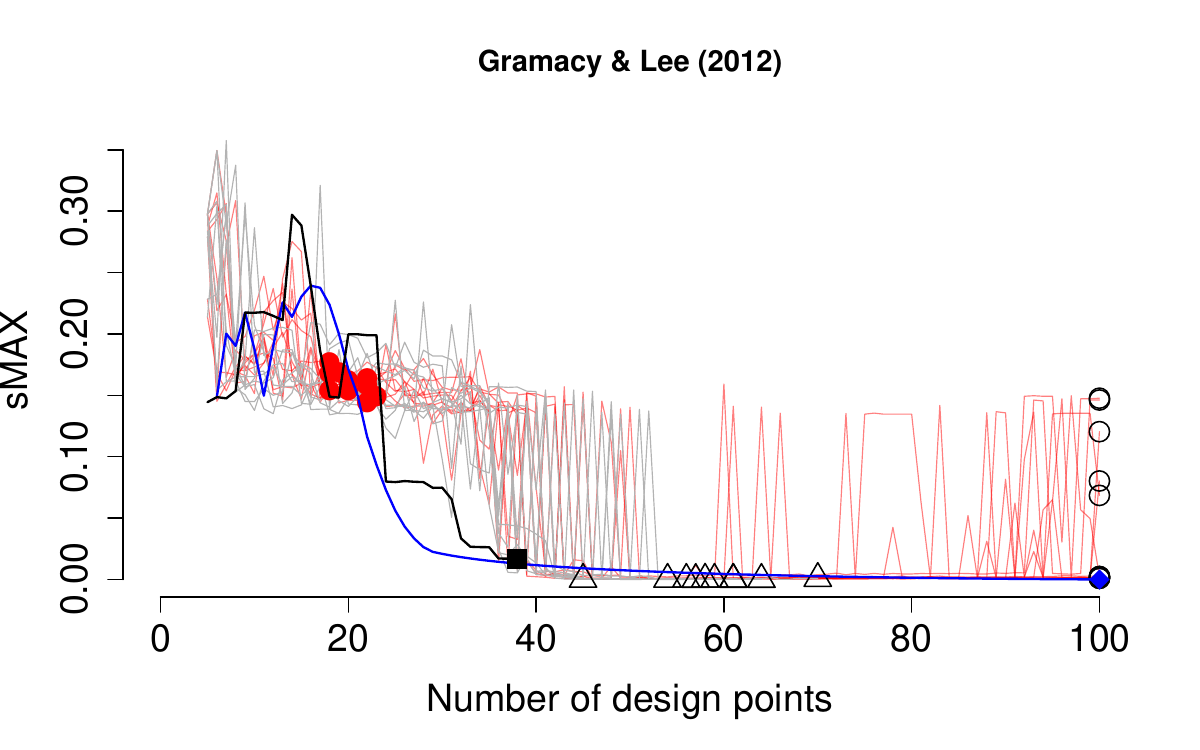}}%
   \parbox[c][1.3in][t]{0.49\textwidth}{\includegraphics[width=.95\hsize, trim=0 0 0 1.8cm, clip]{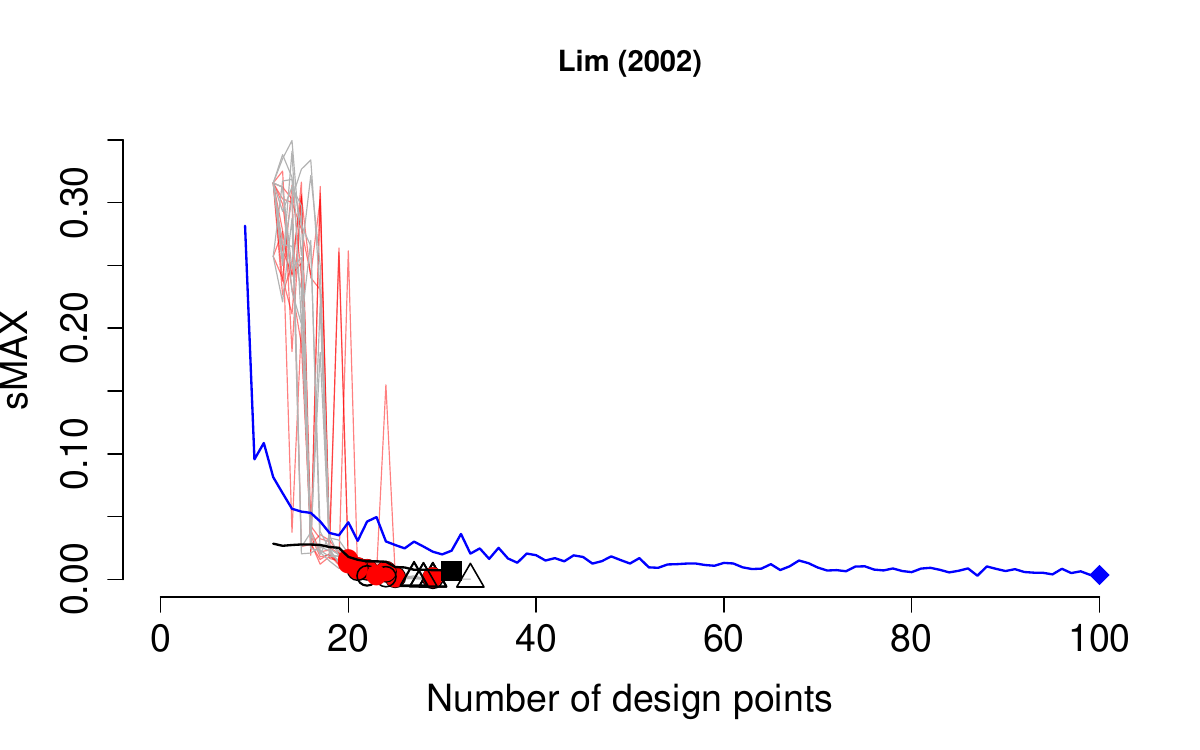}}\\[4ex]
   \parbox[t][.1in][c]{0.49\textwidth}{\subcaption{Gramacy \& Lee (2012)}}
   \parbox[t][.1in][c]{0.49\textwidth}{\subcaption{Lim (2002)}}
   \parbox[c][1.3in][t]{0.49\textwidth}{\includegraphics[width=.95\hsize, trim=0 0 0 2cm, clip]{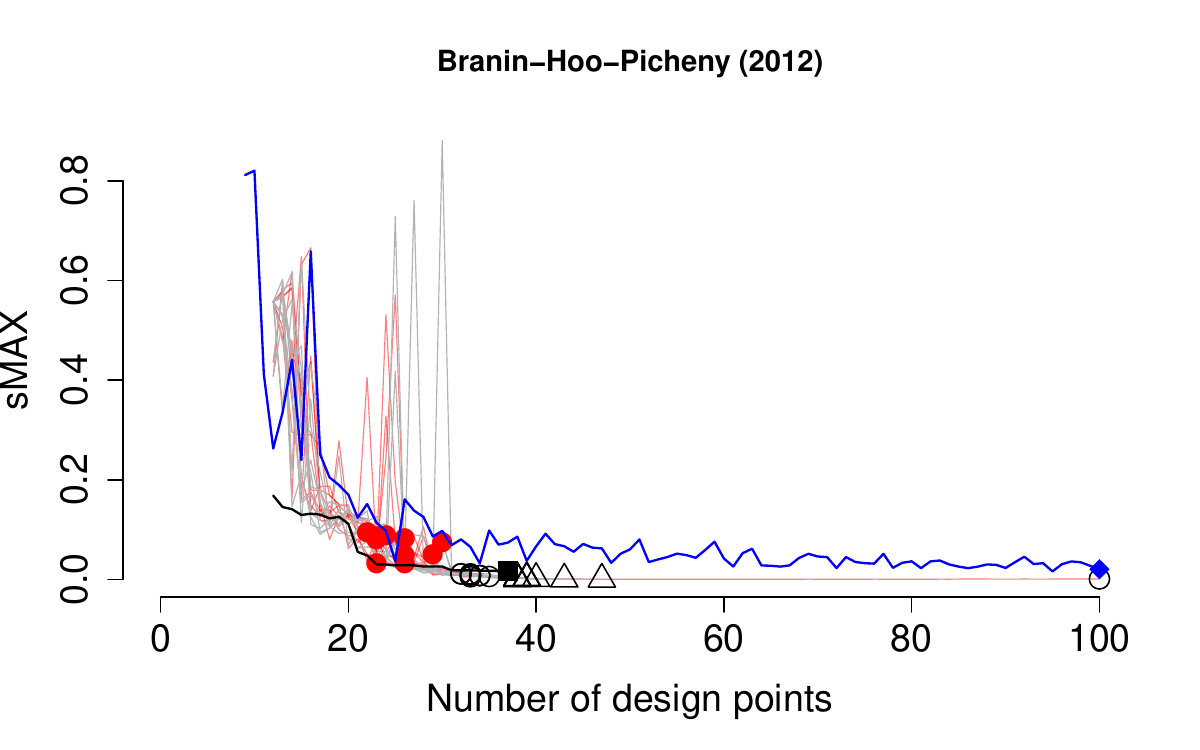}}%
   \parbox[c][1.3in][t]{0.49\textwidth}{\includegraphics[width=.95\hsize, trim=0 0 0 1.8cm, clip]{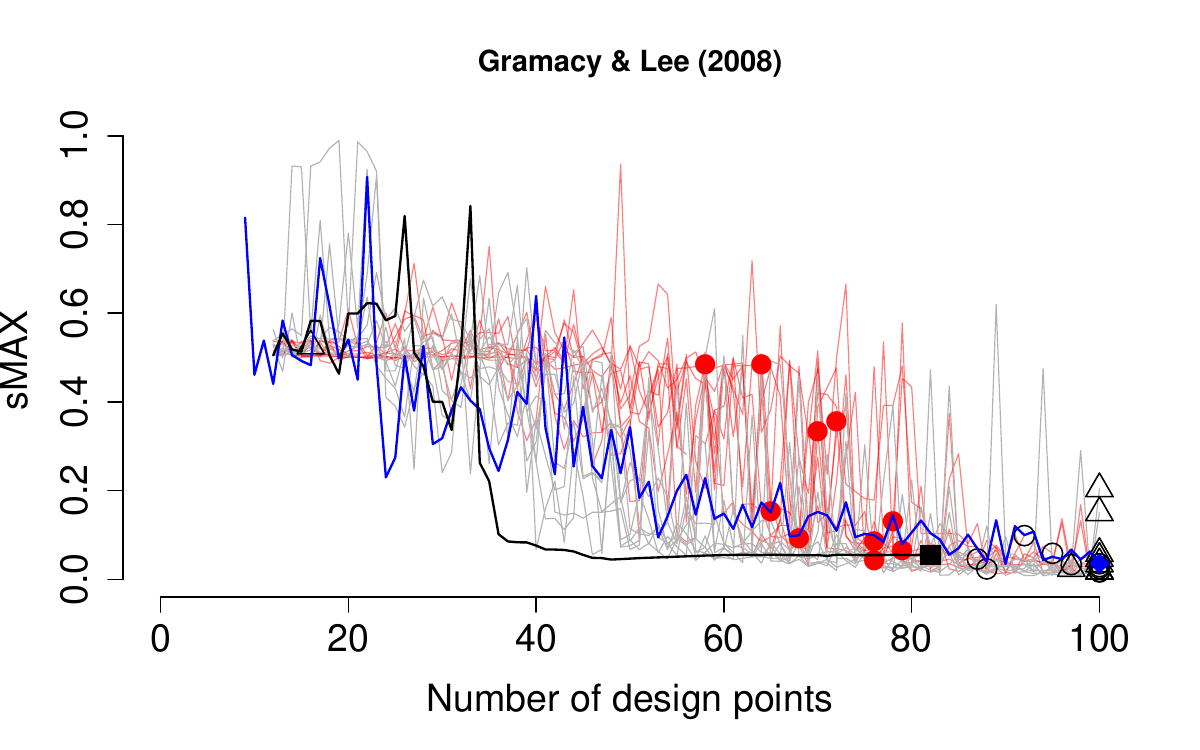}}\\[4ex]
   \parbox[t][.1in][c]{0.49\textwidth}{\subcaption{Branin-Hoo-Picheny (2012)}}
   \parbox[t][.1in][c]{0.49\textwidth}{\subcaption{Gramacy \& Lee (2008)}}
   \parbox[c][1.3in][t]{0.49\textwidth}{\includegraphics[width=.95\hsize, trim=0 0 0 2cm, clip]{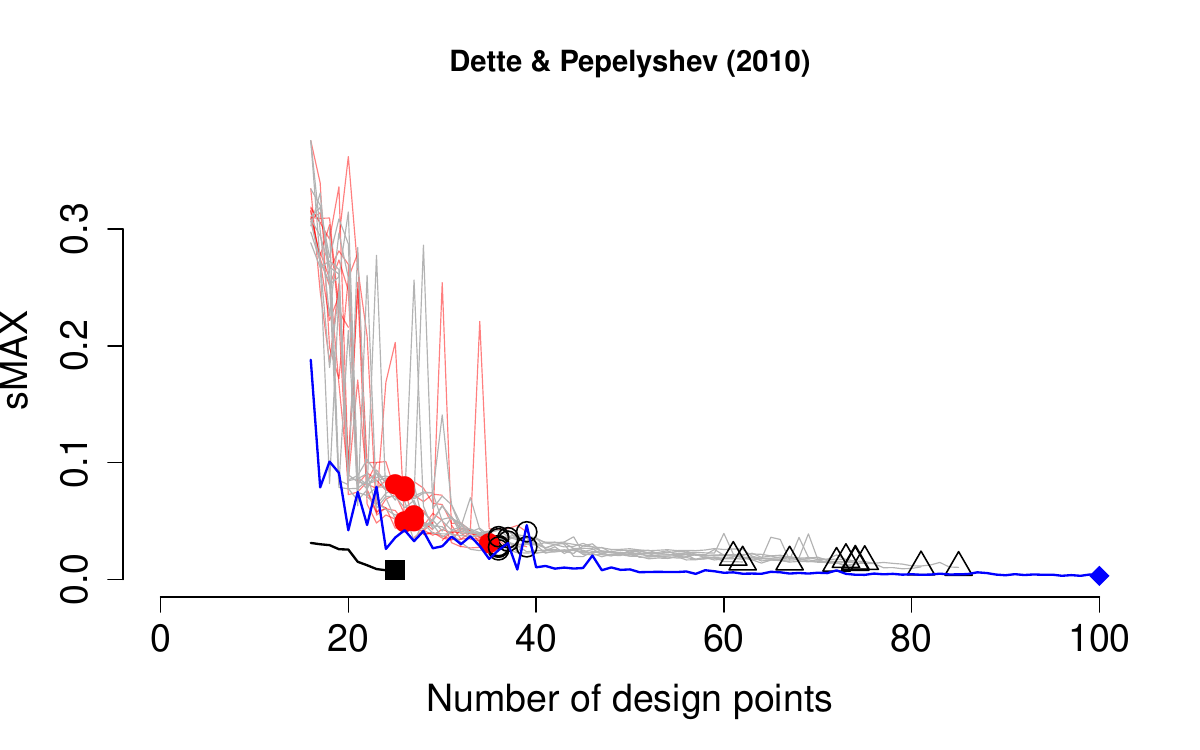}}%
   \parbox[c][1.3in][t]{0.49\textwidth}{\includegraphics[width=.95\hsize, trim=2cm 2cm 2cm 1cm, clip]{Figures/benchmarks/legend.pdf}} \\[4ex]
   \parbox[t][.1in][c]{0.49\textwidth}{\subcaption{Dette \& Pepelyshev (2010)}}
\caption{\label{fig:benchmarks_sMAX}
\textbf{Scaled MAX with various algorithms for the seven benchmark simulators.} ALC | ALM: active learning with the Cohn | MacKay criterion; BTGP: Bayesian treed Gaussian Processes; MaxPro: MaxPro (at the maximum budget).}
\end{figure}

Figure~\ref{fig:benchmarks_sMAX} shows results for sMAX. It has the same layout as Figure~\ref{fig:benchmarks_sRMSE} in the main text, which has results for sRMSE. 

\end{document}